\documentclass[12pt]{article}
\usepackage{graphicx}
\usepackage{amsmath}
\usepackage{bm}
\usepackage{braket}
\usepackage{amsfonts}
\usepackage{amssymb}
\usepackage[colorlinks=true, linkcolor=blue, citecolor=blue, urlcolor=blue]{hyperref}
\usepackage{caption}
\usepackage{subcaption}
\usepackage{geometry}
\usepackage{color}
\usepackage{float}
\usepackage[T1]{fontenc}
\usepackage[utf8]{inputenc}
\usepackage{graphicx}
\usepackage{authblk}
\usepackage{booktabs}
\usepackage[acronym]{glossaries}
\usepackage{siunitx}    
\usepackage[font=small,labelfont=bf]{caption}

\usepackage{amsmath}

\makeatletter
\renewcommand{\[}{\begin{equation}}
\renewcommand{\]}{\end{equation}}
\makeatother

\usepackage[numbers,sort&compress]{natbib} 
\graphicspath{{figures/}{figures/processed/}}

\title{Towards Quantum Advantage in Chemistry}

\glsdisablehyper

\author[1]{Scott N. Genin\thanks{Corresponding author, \href{mailto:scott.genin@otilumionics.com}{scott.genin@otilumionics.com}}}

\author[2]{Ohyun Kwon\thanks{Corresponding author,
\href{mailto:o.kwon@samsung.com}
{o.kwon@samsung.com}}}

\author[1]{Seyyed Mehdi Hosseini Jenab}
\author[2]{Seon-Jeong Lim}
\author[2]{Taehyung Kim}
\author[2]{Tae-Gon Kim}
\author[1]{Rami Gherib}
\author[1]{Angela F. Harper}
\author[1]{Ilya G. Ryabinkin}
\author[1]{Michael G. Helander}

\affil[1]{OTI Lumionics Inc., 3415 American Dr., Mississauga, ON L4V 1T4, Canada}
\affil[2]{Samsung Advanced Institute of Technology (SAIT), Samsung Electronics Co. Ltd., Samsung-ro 130, Yeongtong-gu, Suwon-Si, Gyeonggi-do 16678, Republic of Korea}

\newacronym{MAE}{MAE}{mean absolute error} %

\newacronym{OLED}{OLED}{organic light-emitting diode}     %
\newacronym{NISQ}{NISQ}{noisy intermediate-scale quantum} %
\newacronym{SQD}{SQD}{sample-based quantum diagonalization} %

\newacronym{JW}{JW}{Jordan--Wigner} %
\newacronym{BK}{BK}{Bravyi--Kitaev} %

\newacronym{QPE}{QPE}{quantum phase estimation}          %
\newacronym{VQE}{VQE}{variational quantum eigensolver}   %
\newacronym{QMF}{QMF}{qubit mean-field}                  %
\newacronym{QCC}{QCC}{qubit coupled cluster}             %
\newacronym{iQCC}{iQCC}{iterative qubit coupled cluster} %
\newacronym{DIS}{DIS}{direct interaction set}            %

\newacronym{CAS}{CAS}{complete active space}    %
\newacronym{MO}{MO}{molecular orbital}          %

\newacronym{CI}{CI}{configuration interaction}        %
\newacronym{FCI}{FCI}{full configuration interaction} %
\newacronym{CASCI}{CASCI}{complete active space configuration interaction} %
\newacronym{MCSCF}{MCSCF}{multiconfigurational self-consistent field} %
\newacronym{CASSCF}{CASSCF}{complete active space self-consistent field} %
\newacronym{CC}{CC}{coupled cluster}           %
\newacronym{UCC}{UCC}{unitary coupled cluster} %
\newacronym{UCCSD}{UCCSD}{unitary coupled cluster singles and doubles} %
\newacronym{CCSD}{CCSD}{coupled-cluster singles and doubles} %
\newacronym{CAS-CCSD}{CAS-CCSD}{complete active space coupled-cluster singles and doubles} %
\newacronym{CCSD-T}{CCSD(T)}{coupled-cluster singles and doubles and non-iterative triples} %
\newacronym{CR-CC(2,3)}{CR-CC(2,3)}{completely renormalized coupled-cluster with singles, doubles and perturbative triples correction}
\newacronym{CAS-CR-CC(2,3)}{CAS-CR-CC(2,3)}{complete active space completely renormalized coupled-cluster with singles, doubles and perturbative triples correction}
\newacronym{RHF}{RHF}{restricted Hartree--Fock}                     %
\newacronym{HF}{HF}{Hartree--Fock}                     %
\newacronym{CISD}{CISD}{configuration interaction singles and doubles} %
\newacronym{CAS-CISD}{CAS-CISD}{complete active space configuration interaction singles and doubles} %
\newacronym{ROHF}{ROHF}{restricted open-shell Hartree--Fock}        %
\newacronym{UHF}{UHF}{unrestricted Hartree--Fock}                   %
\newacronym{MPS}{MPS}{matrix product states}                        %
\newacronym{DMRG}{DMRG}{density-matrix renormalization group}       %
\newacronym{DFT}{DFT}{density-functional theory}                    %
\newacronym{TD-DFT}{TD-DFT}{time-dependent density-functional theory} %
\newacronym{PT}{PT}{Epstein-Nesbet perturbation theory}         %

\newacronym{PL}{PL}{photoluminescence}

\date{\today}

\begin{document}
\maketitle


\begin{abstract}
Molecular simulations are widely regarded as leading candidates to demonstrate quantum advantage--defined as the point at which quantum methods surpass classical approaches in either accuracy or scale. Yet the qubit counts and error rates required to realize such an advantage remain uncertain; resource estimates for ground-state electronic structure span orders of magnitude, and no quantum-native method has been validated at a commercially relevant scale. Here we address this uncertainty by executing the iterative qubit coupled-cluster (iQCC) algorithm, designed for fault-tolerant quantum hardware, at unprecedented scale using a quantum solver on classical processors, enabling simulations of transition organo-metallic complexes requiring hundreds of logical qubits and millions of entangling gates. Using this approach, we compute the lowest triplet excited state (T$_1$) energies of Ir(III) and Pt(II) phosphorescent organometallic compounds and show that iQCC achieves the lowest mean absolute error (0.05\,eV) and highest R$^2$ (0.94) relative to experiment, outperforming leading classical methods. We find these systems remain classically tractable up to {\raise.18ex\hbox{$\scriptstyle\sim$}}200 logical qubits, establishing the threshold at which quantum advantage in computational chemistry may emerge and clarifying resource requirements for future quantum computers.

\end{abstract}

\section{Introduction}

Predicting molecular properties with high accuracy remains a central challenge in chemistry, materials, and drug discovery \cite{Marzari2021,Jorgensen2004,Liu2017}. Conventional simulation approaches such as \gls{DFT} are widely used but limited in accuracy \cite{Becke1993,Marzari2021,Kaur2019}, while the gold standard coupled-cluster methods achieve chemical precision only at prohibitive computational cost \cite{manybodymethods}. In principle, quantum computers promise to overcome these barriers by natively representing electronic wavefunctions \cite{Feynman1982,Ladd2010,Cao2019}.
Recent advances have extended the range and now allow classical computers to emulate the \gls{VQE} algorithm in the 80--92 qubit range by using, for example, a sparse representation of the wavefunction~\cite{steiger2025sparsesimulationvqecircuits}, tensor networks on super-computer clusters~\cite{Shang2023,berezutskii2025tensornetworksqc}, or combined hybrid quantum solvers~\cite{RobledoMoreno2025},  but current devices lack the fidelity and scale required for commercially relevant systems \cite{Arute2020,Lee2023,yamamoto2025-qpe-H2-sim}.

Here we present advances in a Quantum Solver that executes the \gls{iQCC} quantum algorithm \cite{Ryabinkin2018,iQCC2020,PT2021} at an unprecedented scale, enabling direct benchmarking of quantum-native approaches against both experiment and state-of-the-art classical methods. By enhancing the efficiency of the \gls{iQCC} Quantum Solver, we simulate problems equivalent to {\raise.17ex\hbox{$\scriptstyle\sim$}}200 logical qubits and $10\times10^{6}$ two-qubit gates, far beyond previous emulations and current quantum hardware \cite{Arute2020,Shang2023,IBMCircuitReduction2024,RobledoMoreno2025,yamamoto2025-qpe-H2-sim}.

As a benchmark, we compute excited state energies of phosphorescent iridium (III) and platinum (II) complexes used in \glspl{OLED} \cite{Lamansky2001,Sajoto2009,Li2017PtII,Li2020ChemMater}. These complexes combine industrial importance with experimentally accessible properties, making them ideal for assessing accuracy. Our results demonstrate systematic improvements over conventional methods and establish a framework for evaluating quantum advantage in chemistry, materials, and drug discovery.
\section{Results}
\subsection{Summary of iQCC}
\begin{figure}
    \centering
    \includegraphics[width=0.99\linewidth]{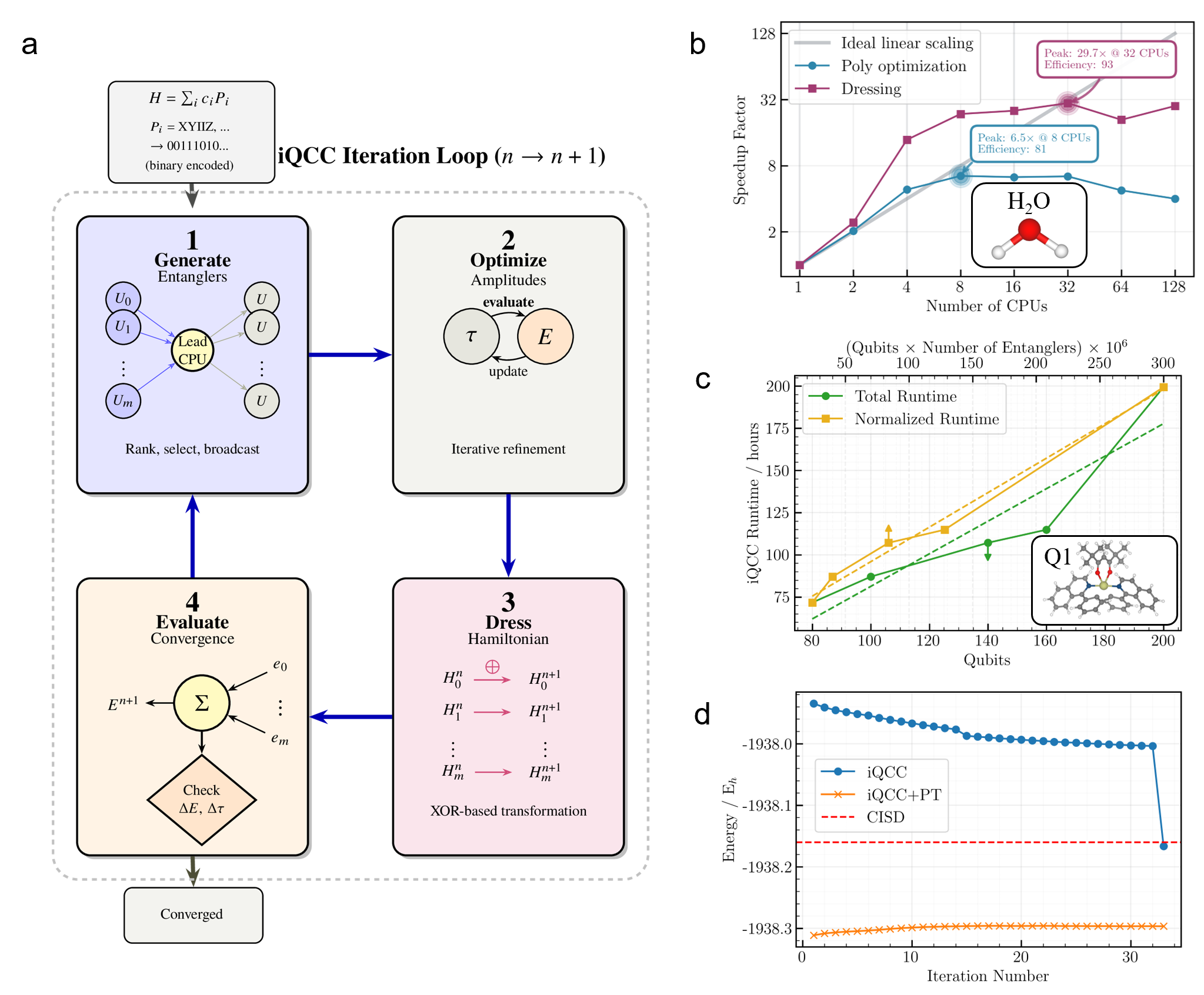}
  \caption{%
  Overview of the computational study for Ir(III) and Pt(II) phosphorescent emitters.
  (a) Schematic representation of the parallel \gls{iQCC} workflow showing CPU-level Hamiltonian partitioning.
  (b) Scaling behavior of the parallel \gls{iQCC} for H$_2$O, and (c) scaling behavior of \gls{iQCC} total runtime for a 200 qubit system, Q1, as a function of both qubits, and normalized qubits $\times$ the number of entanglers required to converge the Hamiltonian as shown in Table \ref{tab:Z}. Dashed lines show a linear fit relative to the scaling data, indicating that with respect to Qubits $\times$ Number of Entanglers, iQCC scales approximately linearly.
  (d) Convergence of the 200 qubit Q1 molecule at CAS(100,100) (200 qubits) with 33 \gls{iQCC} steps converges to the \gls{CISD} value, and the \gls{iQCC}+\gls{PT} surpasses the \gls{CISD} value.}
  \label{fig:study}
\end{figure}

The \gls{iQCC} method is a \gls{VQE} type quantum algorithm which uses a unitary ansatz and represents a trial wave function as
\[
\Psi(\bm{\tau}) = \hat{U}(\bm{\tau})\ket{0},
\]
where $\ket{0}$ is an initial state of a quantum register, typically a one-determinant representation of the ground-state wave function written in a population basis, and $\hat{U}(\bm{\tau})$ is a product of factors $\exp(-i\hat{T}\tau/2)$, where $\hat{T}$ are products of the Pauli matrices. 
This \gls{QCC} form, unlike other commonly used forms, such as the \gls{UCCSD}, derives its generators (``entanglers'') $\hat{T}$ directly from a qubit Hamiltonian to guarantee lowering of the total ground-state energy.
This avoids the barren plateau problem that other implementations struggle with~\cite{Ryabinkin2018,iQCC2020} and does not require amplitudes to be solved using \gls{CCSD} then inputted as an initial guess \cite{Mullinax2025}.
Amplitude optimization is performed using the exact or an approximate ansatz~\cite{Ryabinkin2015} by computing the energy of the Hamiltonian:
\[
E = \bra{\Psi_0} U^\dagger \hat{H} U  \ket{\Psi_0} \,,
\]
where $H = \sum_kc_kP_k$ of which $c_k$ are real coefficients derived from one- and two-electron integrals and $P_k$ are Pauli words. 
In the \gls{iQCC} method, the original Hamiltonian $\hat{H^n}$ is unitary-transformed (``dressed'') using the optimized ansatz $\hat{U}(\bm{\tau})$ into the subsequent iQCC Hamiltonian $\hat{H}^{(n+1)}$(Fig.~\ref{fig:study}a).
At each \gls{iQCC} iteration, a subset of the \gls{DIS} (the set of all possible entanglers with non-zero gradients) is included in the \gls{QCC} ansatz, selected to include the set of all Pauli terms which can be computed efficiently classically. 
To approximately account for the energy contributions of entanglers from the \gls{DIS} that are \textit{not} included into the \gls{QCC} ansatz, we apply the \gls{PT} correction~\cite{PT2021}, to obtain the \gls{iQCC}+\gls{PT} energy estimate.

\subsection{Scalability Analysis and Benchmarking}
\label{sec:results-scaling}
To enable large‑scale classical simulations of \gls{iQCC}, we implemented the algorithm in C\texttt{++} with \texttt{OpenMPI}, parallelizing the most expensive components—operator dressing and amplitude optimization (Fig.~\ref{fig:study}a). The approach uses a bit-partitioning scheme where each Pauli word is represented in symplectic binary form, and selected bits serve as keys that determine CPU assignment. This binary encoding not only yields linear memory scaling, $\mathcal{O}(m*n)$ with respect to the number of qubits and Pauli words $\hat{P}_k$ in the Hamiltonian, but also provides an explicit, computation-free mapping of every Pauli word to its designated processor~\cite{binaryencoding2023} (full details given in Methods Section \ref{sec:methods-parallel-iQCC}).
During the optimization stage, the method supports complete optimization of the exact \gls{QCC} ansatz~\cite{iQCC2020,Genin2022} as well as polynomial approximate forms~\cite{Ryabinkin2015}. Some key advantages for benchmarking chemistry simulations using the \gls{iQCC} quantum solver provides are: 1) full qubit-to-qubit connectivity, 2) the energy can be directly computed without noise giving  consistent and reliable results (see Methods \ref{sec:methods-hamiltonian-construction} for further details), 3) the \gls{QCC} anstaz has a direct circuit equivalent that can be run on other simulators or quantum hardware making it quantum-native \cite{Genin2022,PhysRevA.106.042443,steiger2025sparsesimulationvqecircuits} (details given in Supplementary Information Section \ref{sec:SI4}), and 4) the solver has direct analytical access to the gradient for each parameter $\tau$ in the anstaz. This allows the quantum solver to use derivative based optimization, greatly improving it's efficiency over quantum hardware implementations \cite{Peruzzo_2014_VQE}.
 
The \gls{iQCC} algorithm has several advantages over other quantum simulation methods such as state-vector and tensor networks. 
State-vector methods require storing the full amplitude set $a_z$ in $\ket{\Psi} = \sum_z a_z \ket{z}$, which causes both memory and computational cost to scale as $\mathcal{O}(2^N)$ with respect to the number of qubits \cite{McArdle2020,Weidman2024}.
Tensor network approaches reduce this burden by factorizing $a_z$ into low-rank tensors at cost $\mathcal{O}(N\cdot\chi^p)$, where $\chi$ is the bond dimension and $p$ depends on network topology (typically 3-4), but they remain efficient only when entanglement follows an area law; under volume law entanglement $\chi$ grows rapidly and the cost can approach $\mathcal{O}(2^{N/2})$ or worse \cite{Eisert2014,Biamonte2017,orus2014practical,bridgeman2017handwaving,stoudenmire2012two}. 
Instead of explicitly representing the wavefunction, the \gls{iQCC} algorithm works in the operator space and gives direct access to Hamiltonian coefficients $c_k$ that quantify the contribution of each Pauli term to the energy. 
This distinction carries over to energy evaluation: hardware-based or variational methods estimate $E = \sum_k c_k \left<P_k\right>$ by statistically sampling $\left< P_k \right>$ from repeated measurements, whose sampling requirements grow rapidly with system size. \gls{iQCC} on the other hand, has deterministic access to $c_k$, and their dressed counterparts on a classical computer, permitting exact energy evaluation at each iteration, stable convergence, and precise control over accuracy and computational effort. This efficient parallel \gls{iQCC} implementation allows molecular systems as large as 200 qubits to be simulated with their variational estimate reliably computed below the \gls{CISD} energy using less than 800 GB of RAM (Fig.~\ref{fig:study}c,~\ref{fig:study}d and Table~\ref{tab:Z}).

The strong scaling study of a 36 qubit Hamiltonian of water (\gls{CAS} of 8 electrons and 18 orbitals -- CAS(8,18), $R=0.75$ \AA, basis 6-31G(d) -- shows an excellent speedup from 1 to 32 processes, tapering off beyond 64 processes (Fig.~\ref{fig:study}b). The ansatz optimization steps, which include the construction of the Hessian for amplitude ($\tau$) optimization (Fig.~\ref{fig:study}b blue), benefit less compared to the dressing steps (Fig.~\ref{fig:study}b red). The final energy for water produced by simulating a 2000 entangler \gls{QCC} ansatz was -76.19640996 E$_h$, which is within sub-milliHartree agreement with the literature value of -76.196411 E$_h$ \cite{Ryabinkin2015}.  

In addition to efficient storage and speedup, we further demonstrate that the total wall clock time of the \gls{iQCC} quantum solver is generally linear with respect to qubit count when considering electronic Hamiltonians between systems from 80 to 200 qubits (Fig.~\ref{fig:study}c, Table~\ref{tab:SI-2.2}). 
To illustrate this, we also normalize the scaling reported in Fig.~\ref{fig:study}c with respect to number of final entanglers to be optimized at each qubit count; this demonstrates that the main contributor to the total run time is the non-linear scaling of entangler optimization due to the increase in the required number of parameters to surpass the \gls{CISD} energy.
Considering that other quantum simulations scale as $\mathcal{O}(2^{N/2})$ \cite{Eisert2014,Biamonte2017,orus2014practical,bridgeman2017handwaving,stoudenmire2012two}  or even $\mathcal{O}(2^{N})$ \cite{McArdle2020,Weidman2024}, this performance is outstanding within the scope of the quantum computing simulations currently available.

\begin{table}[htb!]
  \centering
  \caption{\gls{QCC} circuit 2-qubit gate counts and \gls{iQCC} simulation solver times for various \gls{CAS} sizes of Q1 run on two AMD EPYC 7702 64-Core Processors using 64 processes. Gate counts do not include error correction codes and assume a universal quantum computer with full connectivity (i.e. no swap gates required). Hamiltonian terms have a hard limit as described in Section \ref{sec:methods-iqcc}.}
  \label{tab:Z}
  \renewcommand{\arraystretch}{1.1}
  \setlength{\tabcolsep}{4pt} 
  \begin{tabular}{lcccccc}
  \hline
  \addlinespace[3pt]
    \shortstack{Qubit\\Count} & \shortstack{Number of \\ Hamiltonian\\Terms {[10$^9$]}} & \shortstack{Maximum\\Hamiltonian\\Storage {[GB]}} & \shortstack{Number of\\Entanglers} & \shortstack{2-qubit\\CNOT\\Gates {[10$^6$]}} & \shortstack{Variational\\iQCC Energy\\{[E$_h$]}} & \shortstack{CISD\\ Energy\\ {[E$_h$]}} \\
    
    \hline
     80 & 2.48 & 398.2 & 300,000 & 2.859 & -1938.06200 & -1938.05499 \\
     100 & 2.87 & 423.2 & 400,000 & 3.398 & -1938.08067 & -1938.07434 \\
     140 & 2.92 & 537.1 & 600,000 & 4.300 & -1938.11184 & -1938.10940 \\
     160 & 2.99 & 540.0 & 800,000 & 5.597 & -1938.12528 & -1938.12328 \\
     200 & 3.75 & 755.8 & 1,500,000 & 10.210 & -1938.16627 &  -1938.16010   \\
  \hline
  \hline
  \end{tabular}
  \end{table}

Finally, we consider the performance of \gls{iQCC} in the context \gls{VQE}. 
The set of entanglers ($\hat{T}$) optimized by the \gls{iQCC} quantum solver have a direct mapping to quantum circuits (Supplementary Information Section \ref{sec:SI4}). The \gls{iQCC} quantum solver is capable of simulating circuits that have 200 logical qubits (CAS(100,100)) while optimising 1.5$\times10^6$ parameters, which corresponds to an abstract circuit that contains $10.210\times10^6$ 2-qubit CNOT gates (Table \ref{tab:Z}). To date, no universal quantum computer or simulator has executed this number of 2-qubit CNOT gates or qubits for an electronic structure calculation. 
\subsection{Quantum Utility Study: Estimating the peak PL emission of OLED materials}
\begin{figure}
    \centering
    \includegraphics[width=0.7\linewidth]{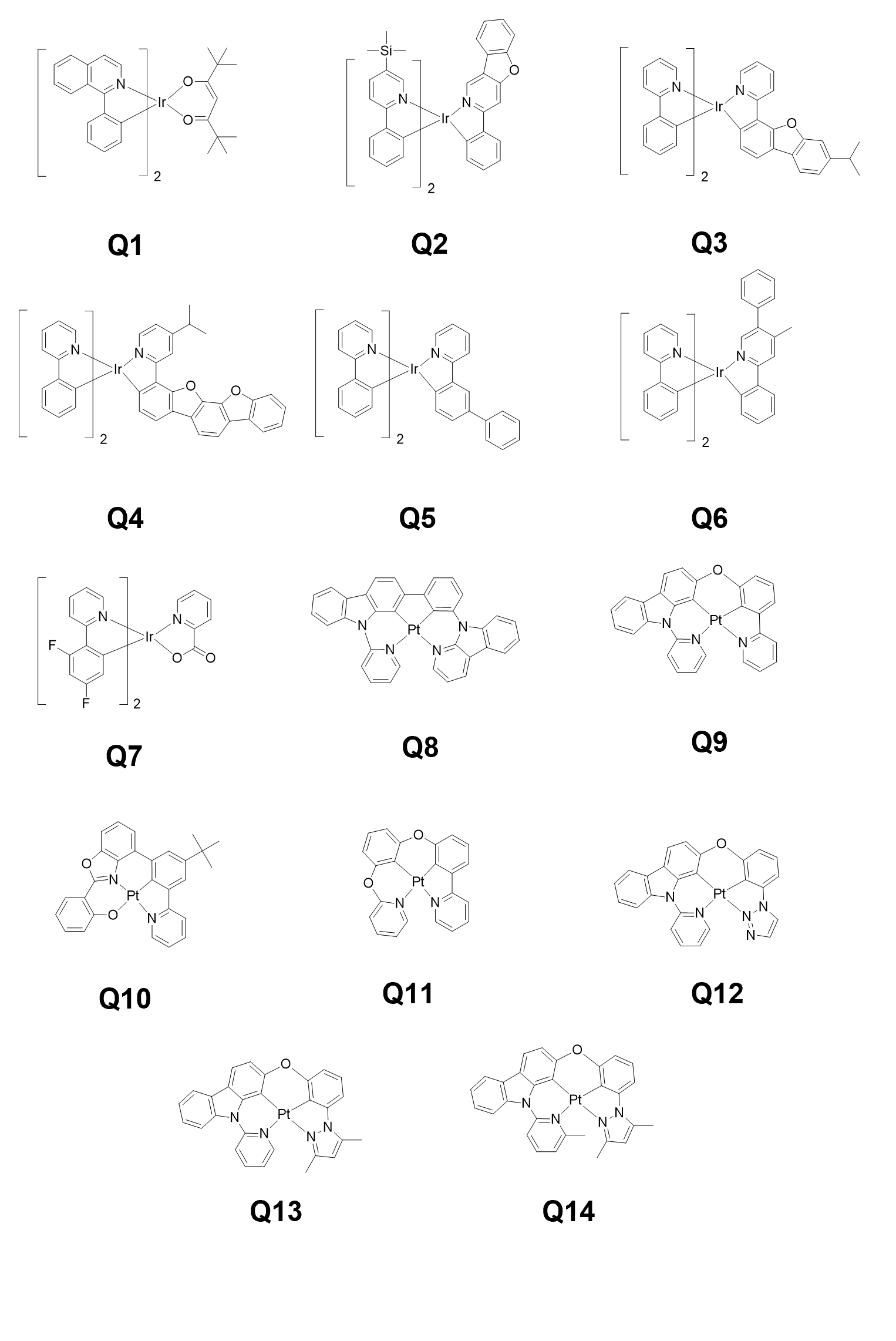}
    \caption{Molecular structures of Ir (III) and Pt (II) complexes used in the benchmarking of \gls{iQCC}. Q1 through Q7 are Ir (III) complexes and Q8 through Q14 are Pt (II) complexes.}
    \label{fig:figure-b}
\end{figure}

To assess the accuracy of the \gls{iQCC} quantum solver in realistic systems, we selected a benchmark set of fourteen Ir (III) and Pt (II) phosphorescent complexes relevant to \gls{OLED}s as shown in Fig.~\ref{fig:figure-b} ~\cite{Li2017PtII,Li2020ChemMater,QS11-1d, QS9,youngminqs13qs16}. Asymmetric Ir (III) complexes were chosen to complement prior studies on symmetric emitters, and the Pt (II) emitters were selected based on well-established scaffolds \cite{Li2017PtII,Li2020ChemMater,youngminqs13qs16}.
These materials span a wide range of phosphorescent emission energies from the deep blue of 440\,nm to orange at 630\,nm. 
 
Experimental \gls{PL} spectra for twelve of the materials were measured at 77\,K (details in Methods Section \ref{sec:methods-synthesis-and-measurement}) using toluene, THF, and dichloromethane which have relatively lower dielectric constants. The \gls{PL} spectra for materials Q12 and Q14 was obtained from literature which were reported measured at 77\,K in 2-MeTHF \cite{Li2017,Li2020ChemMater}.
The low dielectric solvent environment suppresses chromatic shifts, while cryogenic temperatures reduce vibrational broadening \cite{Genin2022}. 
We omit spin orbit coupling since past work on similar compounds found it was not significant \cite{spin-oribit-Ir}. 
Thus, discrepancies between theoretical predictions and experiment are unlikely to arise from solvation or temperature and most likely arise from basis set error, allowing a more direct assessment of the intrinsic accuracy of the quantum and classical methods compared here. 

Within the \gls{OLED} industry, it is quite common to simulate the T$_1 \rightarrow$ S$_0$ gap using either \gls{TD-DFT} or \gls{DFT}, where the B3LYP and CAM-B3LYP functionals have been generally used \cite{Sajoto2009,Li2017,Li2020ChemMater,Genin2022}. 
Past studies on multiple \gls{DFT} functionals demonstrated that long range corrected hybrid functionals with HF exchange correlation, performed the best compared to hybrid functionals \cite{Genin2022}. 
Since we seek to investigate whether quantum advantage can be gained by computing energies of molecular systems, it is important to compare the results against high accuracy methods widely used as benchmarks such as \gls{CCSD} and \gls{CR-CC(2,3)} \cite{piecuch2005breaking,wloch2007extension,robinson2012breaking,schnabel2021limitations,chakraborty2022benchmarking}. After performing the \gls{CR-CC(2,3)} simulations and analyzing the diagnostic amplitude (details in Supplementary Information Section \ref{tab:SI-5.1})\cite{lee1989diagnostic}, it was determined the materials were not strongly correlated, which means that methods such as \gls{DMRG} would not provide a significant advantage over \gls{CR-CC(2,3)}.
These \textit{ab initio} methods have polynomial scaling and have been provided as examples as to why quantum advantage would be difficult to obtain for ground state simulation of weakly to moderately correlated chemical systems \cite{Lee2023}. 

\begin{figure}[htb!]
    \centering
    \includegraphics[width=0.8\linewidth]{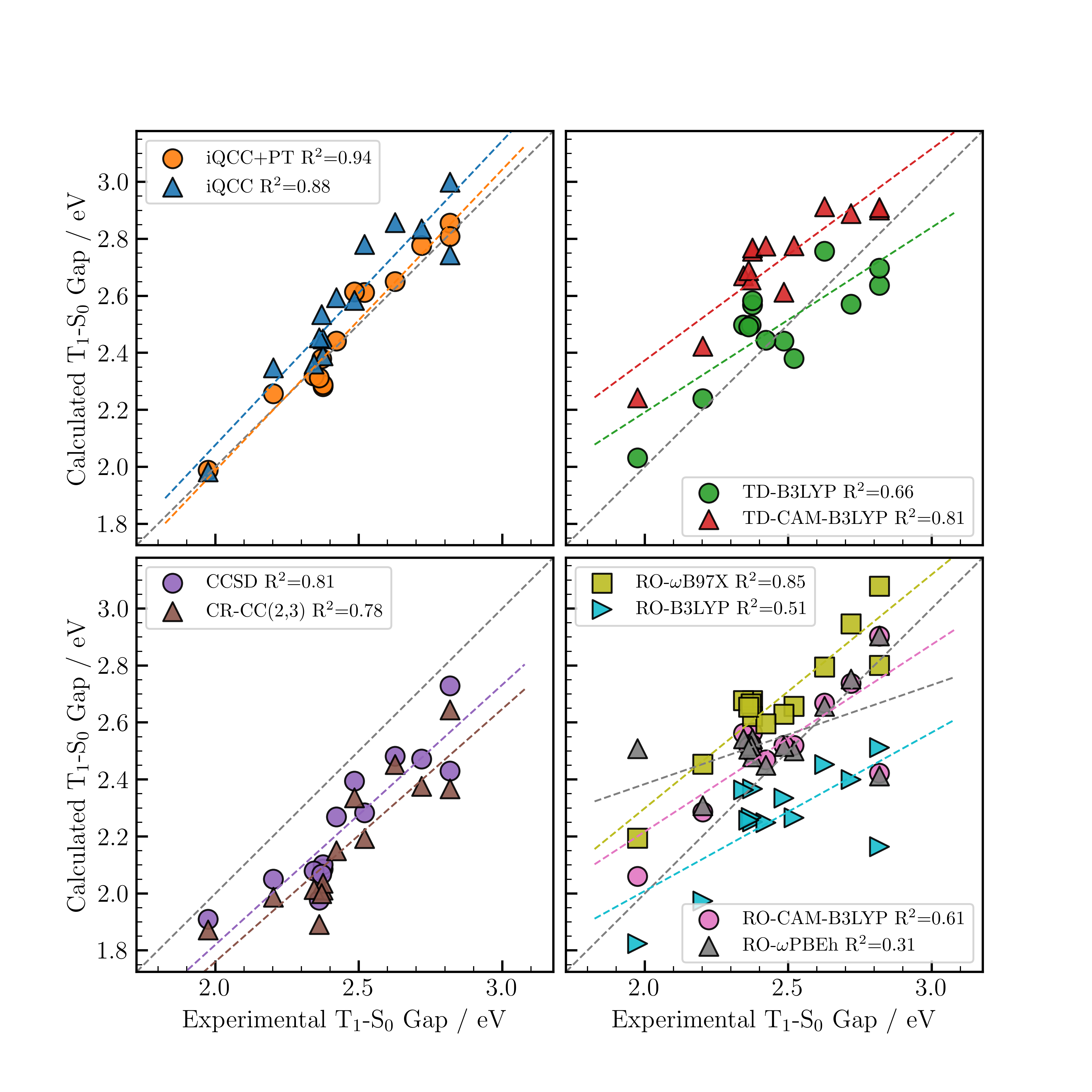} 
    \caption{%
    Absolute values of calculated T$_1$ $\rightarrow$ S$_0$ gap for selected methods from Table \ref{tab:Y} relative to the experimental gap extracted from \gls{PL} spectra at 77\,K. The dashed black line shows a perfect agreement between the experimental and computed gap. \gls{CCSD} and \gls{CR-CC(2,3)} are shown at the bottom left, and exhibit a red shift (lower calculated gap than experiment), whereas several methods \gls{iQCC}, TD-CAM-B3LYP, and RO-$\omega$B97X, all exhibit a blue shift (higher calculated gap than experiment). Results for \gls{HF} are not shown, and are given in Table \ref{tab:Y}.}
    \label{fig:D}
  \end{figure}

A computational chemistry tool needs to be considered precise in order to predict structure-property relationships between changes in chemical structure and changes in electronic structure. 
In an industrial application, the R$^2$ correlation and the P-value are key metrics for evaluating the performance of computational chemistry methods.
Across these materials, both \gls{iQCC} and \gls{iQCC}+\gls{PT} results had the highest performance (Fig.~\ref{fig:D}, and Table~\ref{tab:Y}), based on linear regression metrics R$^2$, P-value, Pearson Correlation, and \gls{MAE}. 
The statistical analysis of the DFT based methods show mixed results, where the best performing DFT functionals based on MAE, tend to perform worse with respect to the linear fit parameters (R$^2$ and P-value shown in Table \ref{tab:Y}).  
The T$_1 \rightarrow$ S$_0$ estimates from \gls{iQCC} alone are less accurate than those from \gls{iQCC}+\gls{PT}, underscoring the need for a \gls{PT} correction unless the full set of entanglers is variationally optimized. 
The ansatz used for the \gls{PT} correction contains the entire list of all entanglers that have a non-zero gradient (which allows for near arbitrary excitation operators that appear as qubit entanglers that can act upon all the qubits in the Hamiltonian) whereas the variationally bounded \gls{iQCC} result is manually limited as to allow the optimization and normalization of the computed energies \cite{PT2021}. 

\begin{figure}[htb!]
    \centering
    \includegraphics[width=0.7\linewidth]{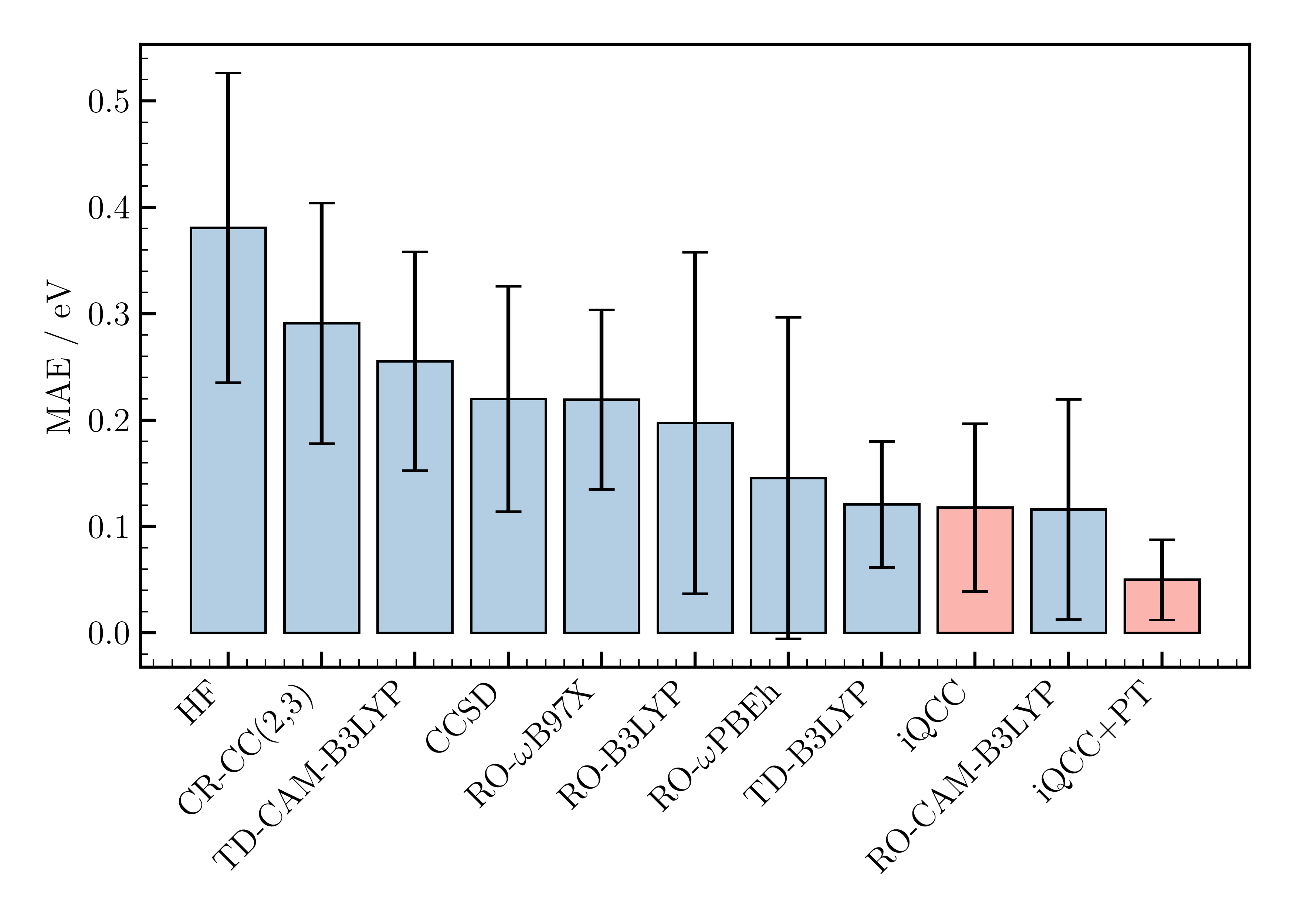}
    \caption{\gls{MAE} of the T$_1$ $\rightarrow$ S$_0$ gap computed using each method referenced in Table \ref{tab:Y} relative to the experimental T$_1$ $\rightarrow$ S$_0$ gap obtained from the \gls{PL} spectrum at 77\,K. Red bars highlight the \gls{iQCC} and \gls{iQCC}+\gls{PT} methods. The \gls{iQCC}+\gls{PT} \gls{MAE} and corresponding standard deviation are lowest of all the methods tested.}
    \label{fig:C}
\end{figure}

\begin{table}[htb!]
  \centering
  \caption{Linear model statistics of the simulation methods against the experimental results, including mean absolute error (MAE), $R^2$ (closer to 1 is better), and P-value (closer to 0 is better).}
  \renewcommand{\arraystretch}{1.2}
  \setlength{\tabcolsep}{6pt} 
  \label{tab:Y}
  \begin{tabular}{lccccc}
  \hline
  \addlinespace[3pt]
  \textbf{Method} & \textbf{$R^2$} & \textbf{P-value} & \textbf{Pearson Corr.} & \textbf{\shortstack{MAE\\{[eV]}}} & \textbf{\shortstack{Std. Dev.\\ {[eV]}}} \\
  \hline
    RHF/ROHF & 0.7393 & 8.04$\times10^{-5}$ & 0.8598 & 0.3809 & 0.1456 \\
    RO-B3LYP & 0.5068 & 4.29$\times10^{-3}$ & 0.7119 & 0.1974 & 0.1604 \\
    RO-CAM-B3LYP & 0.6085 &  9.98$\times10^{-4}$ & 0.7801 & 0.1161 & 0.1035 \\
    RO-$\omega$PBEh & 0.3062 & 4.01$\times10^{-2}$ & 0.5533 & 0.1457 & 0.1512 \\
    RO-$\omega$B97X & 0.8481 & 2.97$\times10^{-6}$ & 0.9209 & 0.2194 & 0.0843 \\
    TD-B3LYP & 0.6638 & 3.86$\times10^{-4}$ & 0.8148 & 0.1209 & 0.0592 \\
    TD-CAM-B3LYP & 0.8103 & 1.15$\times10^{-5}$ & 0.9001 & 0.2555 & 0.1029 \\
    CCSD & 0.8070 & 1.28$\times10^{-5}$ & 0.8983 & 0.2200 & 0.1058 \\
    CR-CC(2,3) & 0.7791 & 2.91$\times10^{-5}$ & 0.8827 & 0.2912 & 0.1130 \\
    iQCC & 0.8786 & 7.61$\times10^{-7}$ & 0.9374 & 0.1180 & 0.0788 \\
    iQCC+PT & 0.9411 & 9.69$\times10^{-9}$ & 0.9701 & 0.0501 & 0.0378 \\
  \hline
  \end{tabular}
  \end{table}

The \gls{iQCC} and \gls{iQCC}+\gls{PT} are generally blue shifted from the experimental value, while \gls{CCSD} and \gls{CR-CC(2,3)} were both consistently red shifted, as shown in Fig.~\ref{fig:D}, which is in agreement with other observations ~\cite{Shen2012}. 
This may indicate that a triplet state is more strongly correlated and \gls{CCSD} and \gls{CR-CC(2,3)} may ``overcorrect'' since they are not, in general, variationally bounded. 
The better \gls{MAE} performance of the \gls{iQCC} and \gls{iQCC}+\gls{PT} compared to the \gls{CCSD} and \gls{CR-CC(2,3)} likely results from the fact that the approximate \gls{QCC} ansatz provides a more balanced description of a wave function by dropping unimportant triple- and disconnected quadruple-excitations while incorporating necessary higher excitation operators. 
The iQCC results are variationally bounded and while the entire \gls{QCC} ansatz was not simulated the \gls{iQCC} results are objectively better than \gls{CCSD} and \gls{CR-CC(2,3)} based on the statistics (Table~\ref{tab:Y}). 
This demonstrates that despite \gls{CR-CC(2,3)} being a common ``gold standard'' benchmark \cite{robinson2012breaking,schnabel2021limitations,chakraborty2022benchmarking}, it would not be a reliable benchmark for quantum algorithms in the field of phosphorescent emitters.  

These results demonstrate that \gls{iQCC}+\gls{PT} not only achieves unprecedented classical emulation scale but also delivers predictive accuracy of phosphorescent T$_1 \rightarrow$ S$_0$ gaps in line with, or surpassing, the best available classical methods. 
By reliably capturing structure–property trends in industrially relevant phosphorescent emitters, \gls{iQCC}+\gls{PT} establishes itself as a practical design tool for next-generation organo-metallic phosphorescent materials and a benchmark for evaluating quantum advantage in computational chemistry. 

\subsection{Implications for Quantum Advantage}
\label{sec:quantum-advantage}

\begin{figure}[htb!]
    \centering
    \includegraphics[width=0.6\linewidth]{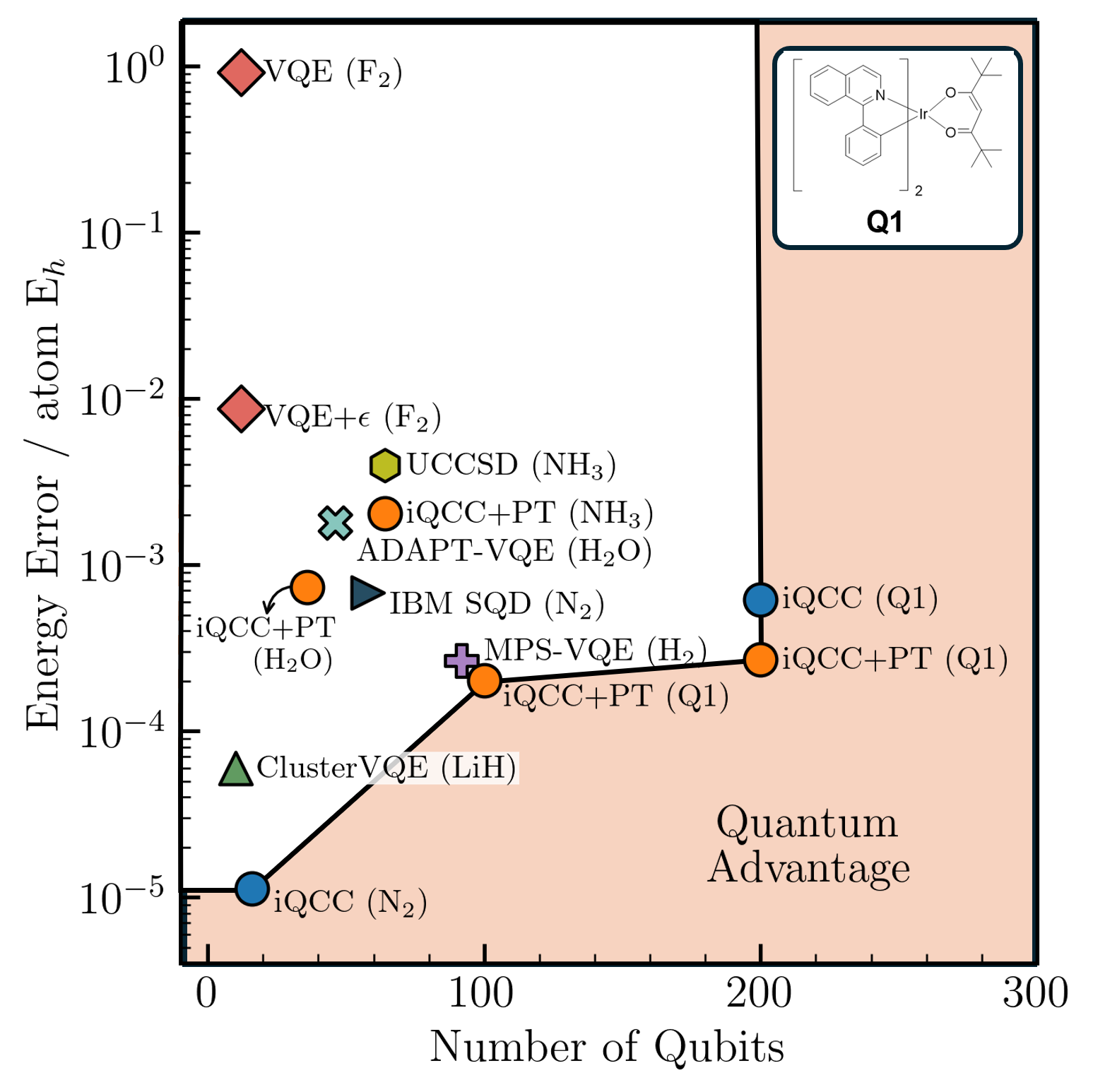}
    \caption{Relative to state of the art \gls{VQE} \cite{guo2024experimental,Shang2023,mullinax2025classical,Mullinax2025,zhang2022variational} \gls{SQD} \cite{robledo2025chemistry} methods, \gls{iQCC} sets a new benchmark for quantum advantage, with respect to accuracy and number of qubits.}
    \label{fig:quantum-advantage}
\end{figure}

If we survey the state-of-the-art simulations which have attempted to benchmark their results against FCI methods, shown in Fig. \ref{fig:quantum-advantage}, we find that \gls{iQCC} outperforms other \gls{VQE}-type methods both in fidelity and qubit scale \cite{zhang2022variational,guo2024experimental,robledo2025chemistry,Mullinax2025,Shang2023} by two to five orders of magnitude without the use of orbital paritioning schemes such as Density Matrix Embedding Theory.
At the 200 qubit level, where \gls{FCI} calculations are impossible, we compare our results for the Q1 molecule to \gls{CISD}+Q results at the same CAS(100,100). To date, no quantum simulator or computer has reached near this number of qubits, and thus our results serve as the new benchmark for ``Quantum Advantage'' in electronic structure calculations. 

 Here we demonstrated the \gls{iQCC} quantum solver, executed on a classical computer, to emulate fault‑tolerant \gls{VQE} for up to 200 logical qubits and $>10\times10^6$ entanglers, and to show that this quantum‑native approach can outperform \gls{CR-CC(2,3)} and DFT/TD‑DFT in predicting T$_1 \rightarrow S_0$ gaps of Ir(III)/Pt(II) complexes for industrial \gls{OLED} applications. The results here show promise for using the \gls{iQCC} in a materials discovery process for new phosphorescent complexes for \gls{OLED}s, specifically Ir (III) and Pt (II) emitters. Additionally, we note that the \gls{iQCC} captures the correct physics by properly attributing the mixed MLCT and LC character in the T$_1$ state. We intend to further push the quantum simulation boundary, by considering both strong correlation and multireference character. Our solver acts as a production tool today, delivering quantum-algorithm results without quantum hardware or expensive supercomputing clusters. Tomorrow, it becomes a benchmarking tool, providing the gold-standard outputs against which quantum computers running chemistry/materials/drug-discovery problems can be validated.
\section{Discussion}

The results here demonstrate that the \gls{iQCC} algorithm if it were to be run on a fully error corrected universal quantum computer would provide a significant advantage in estimating the T$_1 \rightarrow$ S$_0$ gap of Ir (III) and Pt (II) phosphorescent emitters when compared to established classical methods. The correlation plot between the experimental and simulated results for \gls{iQCC} and \gls{iQCC}+\gls{PT} showed R$^2$ of 0.87 and 0.94 respectively, surpassing the best performing conventional methods evaluated. This computational demonstration challenges the conjecture that because there are near linear scaling methods for coupled cluster that can simulate very large molecules, that using a quantum computer to solve for ground state energies would not provide an advantage over conventional methods. 

The results here show promise for using the \gls{iQCC} in a materials discovery process for new \gls{OLED} phosphorescent emitters, specifically Ir (III) and Pt (II) emitters, which are not considered strongly correlated. In future work, we aim to study systems that are known to be strongly correlated and compare against \gls{DMRG} benchmarks. Additionally, we note that the \gls{iQCC} quantum-on-classical-hardware simulations are not limited to the simulation of the T$_1 \rightarrow$ S$_0$ transition, and with the future improvements to the algorithm we intend to further push the quantum simulation boundary, by considering both strong correlation and multireference character.  Ultimately, we aim to position \gls{iQCC} as a competitive tool for quantum chemistry simulations, which provides both efficient and accurate results on a wide range of chemical systems. 

Our solver acts as a production tool today, delivering quantum-algorithm results without quantum hardware or expensive supercomputing clusters. Tomorrow, it becomes a benchmarking tool, providing the gold-standard outputs against which quantum computers running chemistry/materials/drug-discovery problems can be validated.

\section{Methods and Materials}
\label{sec:methods}

\subsection{Qubit Hamiltonian and QCC ansatz}
\label{sec:methods-hamiltonian-construction}
The \gls{QCC} method starts from the second-quantized electronic Hamiltonian of a molecule:
\[
\hat{H} = \sum_{pq} h_{pq} \, a_p^\dagger a_q + \frac{1}{2} \sum_{pqrs} g_{pqrs} \, a_p^\dagger a_q^\dagger a_r a_s,
\]
constructed from fermionic creation and annihilation operators $a_p^\dagger$ and $a_q$. 
The coefficients $h_{pq}$ and $g_{pqrs}$ are values of one- and two-electron integrals written in a spin-orbital basis.

\[
h_{pq} 
= \int \phi_p^{*}(x_1)\, \hat{h}(r_1)\, \phi_q(x_1)\, dx_1 
\]

\[
g_{ijkl} 
= \iint \phi_i^{*}(x_1)\, \phi_j^{*}(x_2)\, 
\frac{1}{r_{12}}\,\phi_k(x_1)\, \phi_l(x_2)\,dx_1\, dx_2 
\]

The electronic Hamiltonian above is constructed using classical methods such as \gls{RHF} or \gls{ROHF} to generate the fermionic Hamiltonian, which is expressed in qubit form as
\begin{equation}
\hat{H} = \sum_{i} c_i \hat{P}_i,
\end{equation}
where each $\hat{P}_i$ is a Pauli word (a tensor product of $X$, $Y$, $Z$, or $I$ operators), and $c_i$ are real coefficients obtained by applying the Jordan--Wigner (JW) transformation to the usual second-quantized fermionic Hamiltonian \cite{JWtransform}. This qubit Hamiltonian forms the basis for the qubit coupled cluster (\gls{QCC}) method.

The QCC ansatz is constructed as a product of one-parameter unitary entangling operations,
\begin{equation}
\hat{U}(\bm{\tau}) = 
\prod_{k=1}^{M} \exp\!\left(-\frac{i}{2} \tau_k \hat{T}_k\right),
\end{equation}
where each $\hat{T}_k$ is a Hermitian Pauli-word generator (an ``entangler''), and each $\tau_k$ is an amplitude to be optimized. The QCC energy for a fixed qubit mean-field reference (in this paper, the Hatree-Fock reference state was used) state $\ket{0}$ is
\begin{equation}
E_{\mathrm{QCC}}(\bm{\tau})
= \bra{0}\hat{U}^\dagger(\bm{\tau})\,\hat{H}\,\hat{U}(\bm{\tau})\ket{0}.
\end{equation}

\subsection{Iterative QCC and generator selection}
\label{sec:methods-iqcc}

The \gls{iQCC} framework constructs the QCC ansatz by adding generators $\hat{T}_\alpha$ that most effectively lower the energy. For a candidate generator, the zero-amplitude energy gradient is
\begin{equation}
g_\alpha
= \left.\frac{\partial E}{\partial \tau_\alpha}\right|_{\tau_\alpha=0}
= -\frac{i}{2}\,\langle 0 |[\hat{H},\hat{T}_\alpha]|0\rangle
= \mathrm{Im}\,\langle 0 | \hat{H}\hat{T}_\alpha | 0 \rangle.
\end{equation}
Generators with large $|g_\alpha|$ provide the steepest local descent directions.

To avoid evaluating gradients for all possible Pauli words, iQCC employs the \acrfull{DIS}, defined as the subset of entanglers guaranteed to yield non-zero gradients. Membership in the DIS is determined by algebraic criteria based on the structure of the Hamiltonian and the action of $\ket{0}\cite{Ryabinkin2018}$.

After optimizing a new amplitude $\tau_{\mathrm{opt}}^{(n)}$ for a generator $\hat{T}$, the Hamiltonian is updated via the dressing transformation as,
\[
\hat{H}^{(n+1)}
= \hat{U}^\dagger(\tau_{\mathrm{opt}}^{(n)})
   \hat{H}^{(n)}
   \hat{U}(\tau_{\mathrm{opt}}^{(n)}) \notag\\
\]
This operation refines the Hamiltonian while preserving its spectrum. The dressing step increases the number of Hamiltonian terms; truncation of small coefficients controls the effective Hamiltonian size.

\paragraph{Outline of the iQCC iteration.}
Each iQCC iteration consists of the following steps:
\begin{enumerate}
\item \textbf{Generate entanglers:} Construct the DIS of Pauli-word generators satisfying the algebraic criteria for non-zero gradients. Evaluate (or estimate) the zero-amplitude gradients $g_\alpha$, rank all DIS generators by $|g_\alpha|$, and select the highest-impact ones to promote to entanglers.
\item \textbf{Optimize amplitudes:} Refine the amplitudes associated with the newly added entanglers within a reduced variational subspace~\cite{Ryabinkin2015}.
\item \textbf{Dress the Hamiltonian:} Apply the unitary dressing transformation using the optimized amplitudes, and truncate Pauli terms whose coefficients fall below the numerical precision threshold.
\item \textbf{Evaluate convergence:} Repeat the iteration until both the energy and amplitude changes satisfy the stopping criteria.
\end{enumerate}

To target specific spin manifolds (e.g., the triplet $T_1$ state), we augment the variational objective with an $\hat{S}^2$ penalty functional~\cite{Ryabinkin2019},
\begin{equation}
E_S(\bm{\tau},\mu)
= E_{\mathrm{QCC}}(\bm{\tau})
+ \mu\left[\bra{0} \hat{S}_{\mathrm{JW}}^2 \ket{0} - S^2 \right]^2,
\end{equation}
where $\hat{S}_{\mathrm{JW}}^2$ is the JW-transformed total spin operator and $\mu$ is a penalty parameter (penalizing either the deviation from number of electrons or spin state).
A full derivation and further details on the QCC ansatz and iQCC procedure from a theoretical perspective are outlined in Ref \cite{iQCC2020}.

\subsection{Parallel iQCC implementation}
\label{sec:methods-parallel-iQCC}
The novelty of the parallel iQCC approach lies in its partitioning strategy: Pauli words are expressed in binary representation, and selected bits are used as partitioning keys \cite{binaryencoding2023}. This provides a clear mapping of where any given Pauli word resides across CPUs. This mapping is particularly critical in the dressing step, where new Pauli words are generated. To maintain efficiency, newly generated terms must be checked against existing ones and combined when appropriate; otherwise, each CPU risks redundantly reconstructing the full Hamiltonian, negating the benefits of parallelization. With the partitioning scheme in place, we can avoid costly all-to-all communication—where every CPU must broadcast its new Pauli words to every other CPU. Since inter-CPU communication is itself a major computational resource, minimizing it is essential. The design of parallel iQCC must therefore satisfy two constraints: (i) distributing Pauli words across CPUs to prevent memory overload on any single node, and (ii) reducing CPU-to-CPU communication to the absolute minimum (saving the time for communication among cpus). Balancing these requirements is key to extending the tractability of iQCC simulations on classical HPC platforms. The partitioning scheme is based on selecting a bitmask that identifies which bits in the binary representation of a Pauli word are used for CPU assignment. The rank of a Pauli word under this mask is computed by packing the selected bits into an integer, which then determines the CPU to which the Pauli word is assigned. This method ensures that each CPU can independently determine the destination of newly generated Pauli words during the dressing step, using simple bitwise operations without requiring extensive communication between CPUs.

A major advantage of this scheme is that Pauli multiplication becomes simple bitwise XOR, allowing each processor to locally determine where newly generated terms belong without global searches or all-to-all communication. This dramatically reduces inter-CPU traffic—one of the dominant costs in large-scale simulations—and distributes memory evenly across nodes. As a result, the parallel C\texttt{++} iQCC implementation achieves significant speedup in dressing and enables efficient optimization of both exact and approximate QCC ansätze at much larger system sizes than previously feasible.

Applying iQCC to large active spaces requires parallelization of both dressing and amplitude optimization. We implemented a distributed-memory parallel version of iQCC in C\texttt{++} with \texttt{OpenMPI}, based on a bit-partitioning strategy for distributing Pauli words across processors.

Each Pauli word $P$ acting on $n$ qubits is encoded in a symplectic binary vector
\[
P \leftrightarrow p = (x_{n-1},\ldots,x_0,\, z_{n-1},\ldots,z_0),
\]
where $x$ and $z$ encode the presence of $X$ and $Z$ operators, respectively. Pauli multiplication corresponds to bitwise XOR on this representation.

A processor assignment is defined using a bitmask $m$ that selects positions of $p$ contributing to the rank,
\begin{equation}
r(P;m) = \sum_{k=0}^{|\mathcal{M}|-1}
         p_{\mathcal{M}_k}\,2^{|\mathcal{M}|-1-k},
\end{equation}
where $\mathcal{M} = \{t : m_t = 1\}$. A Pauli word is mapped to processor 
$\mathrm{cpu}(P)=r(P;m)\bmod N_{\mathrm{cpu}}$.

The essential algebraic property enabling efficient parallelization is
\begin{equation}
r(PQ;m) = r(P;m)\,\oplus\, r(Q;m),
\end{equation}
where $\oplus$ denotes bitwise XOR. Thus, during dressing, each processor can locally determine the destination of any newly generated Pauli word—no all-to-all communication or global lookup tables are required. This eliminates communication bottlenecks, balances memory across nodes, and enables strong scaling of iQCC (see the Results Section \ref{sec:results-scaling}).

\subsection{Perturbative \textit{a posteriori} energy corrections}
\label{sec:PT_energy}

At each iteration of the iQCC procedure, the dressed Hamiltonian $\hat{H}^{(i)}$
defines a direct interaction set (DIS) of generators. Each DIS generator
$\hat{T}_k$ corresponds to a mean-field excited configuration
$\hat{T}_k\ket{0}$ connected to the qubit mean-field reference state
$\ket{0}$. Two quantities required for perturbation theory are computed
during the DIS construction:

\begin{enumerate}
    \item The gradient magnitude
    \begin{equation}
        g_k = \frac{\mathrm{d}E[\hat{T}_k]}{\mathrm{d}\tau}
            = -\frac{\mathrm{i}}{2}
              \braket{0 | [\hat{H}, \hat{T}_k] | 0},
        \label{eq:gradient}
    \end{equation}
    which corresponds to the first energy derivative associated with
    generator $\hat{T}_k$.

    \item The excited-state energy
    \begin{equation}
        E_k = \braket{0 | \hat{T}_k \hat{H} \hat{T}_k | 0}
            = \braket{0 | \hat{X}_k \hat{H} \hat{X}_k | 0},
        \label{eq:excited_energy}
    \end{equation}
    where the second equality uses the factorization
    $\hat{T}_k\ket{0} = \phi_k f_k \hat{X}_k\ket{0}$ to refer to unique
    basis states $\hat{X}_k\ket{0}$.
\end{enumerate}

Defining the reference energy $E_0 = \braket{0|\hat{H}|0}$ and the energy
denominator
\begin{equation}
    D_k = E_0 - E_k,
    \label{eq:denominator}
\end{equation}
the iQCC implementation evaluates two classical \textit{a~posteriori} energy
corrections.

The standard EN2 expression is
\begin{equation}
    \Delta E_{\mathrm{EN2}} = -\sum_k \frac{g_k^2}{D_k},
    \label{eq:EN2}
\end{equation}
yielding the corrected energy
\begin{equation}
    E_{\mathrm{EN2}} = E_0 + \Delta E_{\mathrm{EN2}}.
    \label{eq:EN2_total}
\end{equation}

\subsection{Simulation Protocols}
\label{sec:methods-simulation-protocols}

The 16 qubit nitrogen molecule simulation was conducted at R=2.05 A with cc-pVDZ basis with singlet configuration \cite{Ryabinkin2015}. The 36 qubit water molecule simulation was conducted using equilibrium geometry ($d$(O-H) = 0.95\,\AA\,and $\angle$HOH at 107.6\textdegree) and 6-31G(d) basis \cite{pople-basis}. The \gls{CAS} was generated by using all RHF \glspl{MO} except the lowest-energy MO, which is the core 1s orbital of the oxygen atom, resulting in CAS(8,18) \cite{Ryabinkin2015}. Here, the ground state energy was computed using 2000 entanglers with 2$^{\mathrm{nd}}$ order polynomial \gls{QCC} ansatz.

The singlet and triplet geometries were optimized using either Gaussian~\cite{Gaussian} or TeraChem~\cite{Terachem1,Terachem2} with LANL2DZ-ECP~\cite{Hay1985,Roy2008,TerachemECP} on the metal centers and 6-31G**~\cite{Li2017,Li2020ChemMater,Genin2022,IBMCircuitReduction2024,Ryabinkin2015} on all other atoms, combined with the B3LYP \gls{DFT} functional in the unrestricted formalism~\cite{Becke1993}. \gls{DFT} energies were computed using two methods: \gls{TD-DFT} from the singlet geometry with an S$_0 \rightarrow$ T$_1$ transition employing the B3LYP and CAM-B3LYP functionals~\cite{TDDFT,Becke1993,CAMB3LYP}, and the T$_1 \rightarrow$ S$_0$ energy difference in the triplet geometry (using the triplet optimized geometry for both T$_1$ and S$_0$ electron configuration) with restricted open-shell \gls{DFT} using the B3LYP, CAM-B3LYP, $\omega$PBEh, and $\omega$B97X functionals~\cite{Becke1993,CAMB3LYP}.

The active space used follows the same method described in~\cite{Genin2022} with variable \gls{CAS} sizes. Two Ir complexes (Q2 and Q4) that exhibited poor performance at CAS(70,70) were increased to CAS(100,100) to test whether the \gls{CAS} size was insufficient. Upon completion, it was determined that CAS(100,100) provided much better agreement with experiment compared to CAS(70,70), so the \gls{iQCC}, \gls{CCSD}, and \gls{CR-CC(2,3)} used the larger \gls{CAS} for statistical analysis. To double-check the \gls{iQCC} results, \gls{CCSD} and \gls{CR-CC(2,3)} calculations were also performed using the same one- and two-electron integrals generated by GAMESS~\cite{Schmidt1993,Gordon2005}.

Molecular Hamiltonians in qubit basis were prepared using the triplet geometry of the materials by running triplet \gls{ROHF} and singlet \gls{RHF} to generate the 1 and 2 electron integrals in molecular orbital basis using a modified version of GAMESS \cite{Schmidt1993,Gordon2005}, followed by transformation into qubit basis using the Jordan-Wigner transformation \cite{Nielsen2005}. The selection criteria were to select a fixed number of occupied orbitals below the Fermi level and the same number of virtual orbitals above \cite{Genin2022}. Two Ir complexes Q2 and Q4 which had poor performance at CAS(70,70) were increased to CAS(100,100) to check if the \gls{CAS} was insufficient, and upon finishing the computation it was determined that CAS(100,100) had a much better agreement with experiment compared to CAS(70,70). To double check the \gls{iQCC} results, \gls{CCSD} and \gls{CR-CC(2,3)} was also run using the same 1 \& 2 electron integrals using GAMESS \cite{piecuch2005breaking,wloch2007extension}. \gls{CISD} energies were only run for Q1 to determine the number of entanglers required for the final \gls{iQCC} step. The \gls{CISD} energies for the T$_1$ state at CAS(100,100) could not be converged. 

The \gls{iQCC} algorithm was run using the similar approach to the previous literature \cite{Ryabinkin2019,Genin2022,Ryabinkin2015}. In this instance, there were 8 \gls{iQCC} dressing steps that contained 28 entanglers of which the entanglers were optimized to the 6th order, followed by 6 \gls{iQCC} steps with 46 entanglers optimized at the 4th order. This was then followed by a repeating procedure that would optimize 300 entanglers at the 2nd order, but only dress 46 entanglers per \gls{iQCC} step until the amplitudes of all 300 entanglers was less than 0.012 at which a final \gls{iQCC} step was conducted using a set number of entanglers at the 1st order depending on the \gls{CAS} size (300,000 entanglers for CAS(40,40), 400,000 entanglers for CAS(50,50), 600,000 entanglers for CAS(70,70), 800,000 entanglers for CAS(80,80), and 1,500,000 entanglers for CAS(100,100)). The \gls{iQCC} simulations for the \gls{OLED} phosphorescent emitters were run on compute nodes using two AMD EPYC 7702 64-core processors with 64 processes. 

\subsection{Synthesis and Measurement}
\label{sec:methods-synthesis-and-measurement}
Seven Ir (III) and five Pt (II) phosphorescent materials were either synthesized using published synthetic routes \cite{Li2017PtII,Li2020ChemMater,QS11-1d,QS9,youngminqs13qs16} or purchased. The PL spectra were measured using Hamamatsu Quantaurus-QY C11347 spectrometer at 77\,K. Materials Q1, Q5, Q9, Q13, Q6, Q10, Q3, Q4, Q2 were measured in toluene, Q7 was measured in tetrahydrofuran, and Q8 and Q11 were measured in dichloromethane. The literature values for the peak PL emission of Q12 and Q14 were reported to be measured in 2-methyltetrahydrofuran at 77\,K \cite{Li2017PtII,Li2020ChemMater}.

\section{Acknowledgments}
The authors would like to acknowledge the funding provided by Innovation Solutions Canada project \#202208-F0033-C00003 and Next Generation Manufacturing Grant \#14234. The authors would like to thank Hyun-Ju Lim at the Embassy of Canada in the Republic of Korea for helping facilitate the collaboration through the Canadian International Innovation Program.

\bibliographystyle{unsrt}
\bibliography{bibliography/references}

\begin{thebibliography}{10}

\bibitem{Marzari2021}
Nicola Marzari, Andrea Ferretti, and Chris Wolverton.
\newblock Electronic-structure methods for materials design.
\newblock {\em Nature Materials}, 20:736--749, 2021.

\bibitem{Jorgensen2004}
William~L. Jorgensen.
\newblock The many roles of computation in drug discovery.
\newblock {\em Science}, 303:1813--1818, 2004.

\bibitem{Liu2017}
Hanyu Liu, Ivan~I. Naumov, Roald Hoffmann, N.~W. Ashcroft, and Russell~J.
  Hemley.
\newblock Potential high-$t_c$ superconducting lanthanum and yttrium hydrides
  at high pressure.
\newblock {\em Proceedings of the National Academy of Sciences},
  114:6990--6995, 2017.

\bibitem{Becke1993}
Axel~D. Becke.
\newblock Density-functional thermochemistry. {III}. {T}he role of exact
  exchange.
\newblock {\em Journal of Chemical Physics}, 98(7):5648--5652, 1993.

\bibitem{Kaur2019}
Jaspreet Kaur, Fabian Egor, and Viktor~N. Staroverov.
\newblock What is the accuracy limit of adiabatic linear-response {TDDFT} using
  exact exchange–correlation potentials and approximate kernels?
\newblock {\em Journal of Chemical Theory and Computation}, 15(7):3665--3677,
  2019.

\bibitem{manybodymethods}
Isaiah Shavitt and Rodney~J. Bartlett.
\newblock {\em Many-body Methods in Phyiscs and Chemistry}.
\newblock Cambridge University Press, 2009.

\bibitem{Feynman1982}
Richard~P. Feynman.
\newblock Simulating physics with computers.
\newblock {\em International Journal of Theoretical Physics}, 21(6-7):467--488,
  1982.

\bibitem{Ladd2010}
T.~D. Ladd, Fedor Jelezko, Raymond Laflamme, Yasunobu Nakamura, Christopher
  Monroe, and Jeremy~L. O’Brien.
\newblock Quantum computers.
\newblock {\em Nature}, 464(7285):45--53, 2010.

\bibitem{Cao2019}
Yudong Cao, Jonathan Romero, Jonathan~P. Olson, Matthias Degroote, Peter~D.
  Johnson, M{\'a}ria Kieferov{\'a}, Ian~D. Kivlichan, Tim Menke, Borja
  Peropadre, Nicolas P.~D. Sawaya, Sukin Sim, Libor Veis, and Al{\'a}n
  Aspuru-Guzik.
\newblock Quantum chemistry in the age of quantum computing.
\newblock {\em Chemical Reviews}, 119:10856--10915, 2019.

\bibitem{steiger2025sparsesimulationvqecircuits}
Damian~S. Steiger, Thomas H\"{a}ner, Scott~N. Genin, and Helmut~G. Katzgraber.
\newblock Sparse simulation of {VQE} circuits.
\newblock {\em arXiv preprint \emph{arXiv:2404.10047}}, 2025.

\bibitem{Shang2023}
Hongxiang Shang, Yuxiang Fan, Li~Shen, Qiming Chen, Xiaohu Qian, Chao Yang,
  Ying Li, Xiao Yuan, Yu~Wang, Lei Zhang, Yonggang Yao, Xiaoliang Yao, Junning
  Sun, Mingcong Qin, Xiaoming Zhou, Jian Fang, Xiaotong Gong, Jian Zhou, Tian
  Luo, Linfeng Yuan, Bing Xiao, Yujun Zheng, Guang Chen, Xin Liu, Lu~Xu, Zewen
  Ma, Xiaobo Hu, Yong Liu, Guangming Chen, Wei Xue, Ying Xu, Guojing Chen,
  Jianfeng Lu, Guanghao Chen, Yinan He, Xiaowei Li, et~al.
\newblock Towards practical and massively parallel quantum computing emulation
  for quantum chemistry.
\newblock {\em npj Quantum Information}, 9:33, 2023.

\bibitem{berezutskii2025tensornetworksqc}
Aleksandr Berezutskii, Minzhao Liu, Atithi Acharya, Roman Ellerbrock, Johnnie
  Gray, Reza Haghshenas, Zichang He, Abid Khan, Viacheslav Kuzmin, Dmitry
  Lyakh, Danylo Lykov, Salvatore Mandrà, Christopher Mansell, Alexey Melnikov,
  Artem Melnikov, Vladimir Mironov, Dmitry Morozov, Florian Neukart, Alberto
  Nocera, Michael~A. Perlin, Michael Perelshtein, Matthew Steinberg, Ruslan
  Shaydulin, Benjamin Villalonga, Markus Pflitsch, Marco Pistoia, Valerii
  Vinokur, and Yuri Alexeev.
\newblock Tensor networks for quantum computing.
\newblock {\em arXiv preprint \emph{arXiv:2503.08626}}, 2025.

\bibitem{RobledoMoreno2025}
Javier Robledo-Moreno, Mario Motta, Holger Haas, Ali Javadi-Abhari, et~al.
\newblock Chemistry beyond the scale of exact diagonalization on a
  quantum-centric supercomputer.
\newblock {\em Science Advances}, 11(25):eadu9991, 2025.

\bibitem{Arute2020}
Frank Arute, Kunal Arya, Ryan Babbush, Dave Bacon, Joseph~C. Bardin, Rami
  Barends, Jacopo Basso, et~al.
\newblock Hartree-{F}ock on a superconducting qubit quantum computer.
\newblock {\em Science}, 369:1084--1089, 2020.

\bibitem{Lee2023}
Seunghoon Lee, Joonho Lee, Huanchen Zhai, Yu~Tong, Alexander~M. Dalzell,
  Ashutosh Kumar, Phillip Helms, Johnnie Gray, Zhi-Hao Cui, Wenyuan Liu,
  Michael Kastoryano, Ryan Babbush, John Preskill, David~R. Reichman, Earl~T.
  Campbell, Edward~F. Valeev, Lin Lin, and Garnet Kin-Lic Chan.
\newblock Evaluating the evidence for exponential quantum advantage in
  ground‐state quantum chemistry.
\newblock {\em Nature Communications}, 14(1):1952, 2023.

\bibitem{yamamoto2025-qpe-H2-sim}
Kentaro Yamamoto, Yuta Kikuchi, David Amaro, Ben Criger, Silas Dilkes, Ciarán
  Ryan-Anderson, Andrew Tranter, Joan~M. Dreiling, Dan Gresh, Cameron Foltz,
  Michael Mills, Steven~A. Moses, Peter~E. Siegfried, Maxwell~D. Urmey,
  Justin~J. Burau, Aaron Hankin, Dominic Lucchetti, John~P. Gaebler, Natalie~C.
  Brown, Brian Neyenhuis, and David~Muñoz Ramo.
\newblock Quantum error-corrected computation of molecular energies.
\newblock {\em arXiv preprint: \emph{arXiv:2505.09133}}, 2025.

\bibitem{Ryabinkin2018}
Ilya~G. Ryabinkin, Tzu-Ching Yen, Scott~N. Genin, and Artur~F. Izmaylov.
\newblock Qubit {C}oupled {C}luster method: A systematic approach to quantum
  chemistry on a quantum computer.
\newblock {\em Journal of Chemical Theory and Computation}, 14(12):6317--6326,
  2018.

\bibitem{iQCC2020}
Ilya~G. Ryabinkin, Robert~A. Lang, Scott~N. Genin, and Artur~F. Izmaylov.
\newblock Iterative qubit coupled cluster approach with efficient screening of
  generators.
\newblock {\em Journal of Chemical Theory and Computation}, 16(2):1055--1063,
  2020.

\bibitem{PT2021}
Ilya~G. Ryabinkin, Artur~F. Izmaylov, and Scott~N. Genin.
\newblock A posteriori corrections to the iterative qubit coupled cluster
  method to minimize the use of quantum resources in large-scale calculations.
\newblock {\em Quantum Science and Technology}, 6(2):024012, 2021.

\bibitem{IBMCircuitReduction2024}
Partha Suryanarayanan, Shreyans Sethi, and et~al.
\newblock Increasing the reach of quantum computing algorithms for chemistry
  applications using novel circuit reduction and {Z2} symmetries-based methods,
  2024.
\newblock Presented at ACS Fall 2024.

\bibitem{Lamansky2001}
Stanley Lamansky, Peter Djurovich, David Murphy, Faiz Abdel-Razzaq, Hong-Eun
  Lee, Chihaya Adachi, Paul~E. Burrows, Stephen~R. Forrest, and Mark~E.
  Thompson.
\newblock Highly phosphorescent bis-cyclometalated iridium complexes:
  synthesis, photophysical characterization, and use in organic light emitting
  diodes.
\newblock {\em Journal of the American Chemical Society}, 123(18):4304--4312,
  2001.

\bibitem{Sajoto2009}
Timmy Sajoto, Peter~I. Djurovich, Adrian~B. Tamayo, Jonas Oxgaard, William~A.
  Goddard, and Mark~E. Thompson.
\newblock Blue and near-{UV} phosphorescence from iridium complexes with
  cyclometalated tetraarylimidazole ligands.
\newblock {\em Journal of the American Chemical Society}, 131(27):9813--9822,
  2009.

\bibitem{Li2017PtII}
Guijie Li, Alicia Wolfe, Jason Brooks, Zhi-Qiang Zhu, and Jian Li.
\newblock Modifying emission spectral bandwidth of phosphorescent
  platinum({II}) complexes through synthetic control.
\newblock {\em Inorganic Chemistry}, 56(14):8244--8256, 2017.

\bibitem{Li2020ChemMater}
Guijie Li, Xiangdong Zhao, Tyler Fleetham, Qidong Chen, Feng Zhan, Jianbing
  Zheng, Yun-Fang Yang, Weiwei Lou, Yuning Yang, Kun Fang, Zongzhou Shao,
  Qisheng Zhang, and Yuanbin She.
\newblock Tetradentate {P}latinum({II}) complexes for highly efficient
  phosphorescent emitters and sky blue oleds.
\newblock {\em Chemistry of Materials}, 32(1):537--548, 2020.

\bibitem{Mullinax2025}
J.~Wayne Mullinax and Norm~M. Tubman.
\newblock Large-scale sparse wave function circuit simulator for applications
  with the variational quantum eigensolver.
\newblock {\em The Journal of Chemical Physics}, 162(7):074114, 02 2025.

\bibitem{Ryabinkin2015}
Ilya~G. Ryabinkin, Seyyed~Mehdi Hosseini~Jenab, and Scott~N. Genin.
\newblock Optimization of the qubit coupled cluster ansatz on classical
  computers.
\newblock {\em Journal of Chemical Theory and Computation}, 11(8):3616--3625,
  2025.

\bibitem{binaryencoding2023}
Artur Izmaylov, Robert~A. Lang, Scott~N. Genin, and Ilya Ryabinkin.
\newblock Methods and systems for solving a problem using qubit coupled cluster
  and binary encoding of quantum information, 2023.
\newblock Canadian Patent 3238140A1,10 Nov 2022, Assignee: OTI Lumionics Inc.,
  Country: Canada.

\bibitem{Genin2022}
Scott~N. Genin, Ilya~G. Ryabinkin, Nathan~R. Paisley, Sarah~O. Whelan,
  Michael~G. Helander, and Zachary~M. Hudson.
\newblock Estimating phosphorescent emission energies in {Ir(III)} complexes
  using large-scale quantum computing simulations.
\newblock {\em Angewandte Chemie International Edition}, 61:e202116175, 2022.

\bibitem{PhysRevA.106.042443}
Andrew Jena, Scott~N. Genin, and Michele Mosca.
\newblock Optimization of variational-quantum-eigensolver measurement by
  partitioning pauli operators using multiqubit clifford gates on noisy
  intermediate-scale quantum hardware.
\newblock {\em Phys. Rev. A}, 106:042443, Oct 2022.

\bibitem{Peruzzo_2014_VQE}
Alberto Peruzzo, Jarrod McClean, Peter Shadbolt, Man-Hong Yung, Xiao-Qi Zhou,
  Peter~J. Love, Alán Aspuru-Guzik, and Jeremy~L. O’Brien.
\newblock A variational eigenvalue solver on a photonic quantum processor.
\newblock {\em Nature Communications}, 5(1), July 2014.

\bibitem{McArdle2020}
Sam McArdle, Suguru Endo, Al{\'a}n Aspuru-Guzik, Simon~C. Benjamin, and Xiao
  Yuan.
\newblock Quantum computational chemistry.
\newblock {\em Reviews of Modern Physics}, 92(1):015003, 2020.

\bibitem{Weidman2024}
Jared~D Weidman, Manas Sajjan, Camille Mikolas, Zachary~J Stewart, Johannes
  Pollanen, Sabre Kais, and Angela~K Wilson.
\newblock Quantum computing and chemistry.
\newblock {\em Cell Reports Physical Science}, 5(9), 2024.

\bibitem{Eisert2014}
Yimin Ge and Jens Eisert.
\newblock Area laws and efficient descriptions of quantum many-body states.
\newblock {\em New Journal of Physics}, 18(8):083026, 2016.

\bibitem{Biamonte2017}
Jacob Biamonte and Ville Bergholm.
\newblock Tensor networks in a nutshell.
\newblock {\em arXiv preprint \emph{arXiv:1708.00006}}, 2017.

\bibitem{orus2014practical}
Rom{\'a}n Or{\'u}s.
\newblock A practical introduction to tensor networks: Matrix product states
  and projected entangled pair states.
\newblock {\em Annals of Physics}, 349:117--158, 2014.

\bibitem{bridgeman2017handwaving}
Jacob~C Bridgeman and Christopher~T Chubb.
\newblock Hand-waving and interpretive dance: an introductory course on tensor
  networks.
\newblock {\em Journal of Physics A: Mathematical and Theoretical},
  50(22):223001, 2017.

\bibitem{stoudenmire2012two}
Edwin~M Stoudenmire and Steven~R White.
\newblock Studying two-dimensional systems with the density matrix
  renormalization group.
\newblock {\em Annual Review of Condensed Matter Physics}, 3(1):111--128, 2012.

\bibitem{QS11-1d}
Guangzhao Lu, Shuaibing Li, Yangke Long, Yang Li, Changjiang Zhou, Xinzhong
  Wang, and Liang Zhou.
\newblock Asymmetric tris-heteroleptic iridium({III}) complexes towards blue
  phosphorescence: Synthesis, photophysics and {OLED} application.
\newblock {\em Dyes and Pigments}, 220:111713, 2023.

\bibitem{QS9}
Y.~Zhang, X.~Chen, J.~Wang, et~al.
\newblock Enhanced emitting dipole orientation based on asymmetric
  iridium({III}) complexes.
\newblock {\em Advanced Science}, 11(xx):e202402349, 2024.
\newblock Asymmetric Ir(III) design improving EDO and outcoupling.

\bibitem{youngminqs13qs16}
Youngmin You.
\newblock Pt({II}) complexes with tetradentate ligands: Toward commercially
  applicable blue organic electroluminescence devices.
\newblock {\em Coordination Chemistry Reviews}, 526:216374, 2025.

\bibitem{Li2017}
Guijie Li, Kody Klimes, Tyler Fleetham, Zhi-Qiang Zhu, and Jian Li.
\newblock Stable and efficient sky-blue organic light emitting diodes employing
  a tetradentate platinum complex.
\newblock {\em Applied Physics Letters}, 110(11):113301, 2017.

\bibitem{spin-oribit-Ir}
Arthur R.~G. Smith, Paul~L. Burn, and Ben~J. Powell.
\newblock Spin–orbit coupling in phosphorescent iridium({III}) complexes.
\newblock {\em ChemPhysChem}, 12(13):2429--2438, 2011.

\bibitem{piecuch2005breaking}
P~Piecuch and M~Wloch.
\newblock Breaking bonds with the left eigenstate completely renormalized
  coupled-cluster method.
\newblock {\em Journal of Chemical Physics}, 123(224105):1--10, 2005.

\bibitem{wloch2007extension}
Marta W{\l}och, Jeffrey~R Gour, and Piotr Piecuch.
\newblock Extension of the renormalized coupled-cluster methods exploiting left
  eigenstates of the similarity-transformed hamiltonian to open-shell systems:
  A benchmark study.
\newblock {\em The Journal of Physical Chemistry A}, 111(44):11359--11382,
  2007.

\bibitem{robinson2012breaking}
James~B Robinson and Peter~J Knowles.
\newblock Breaking multiple covalent bonds with hartree--fock-based quantum
  chemistry: quasi-variational coupled cluster theory with perturbative
  treatment of triple excitations.
\newblock {\em Physical Chemistry Chemical Physics}, 14(19):6729--6732, 2012.

\bibitem{schnabel2021limitations}
Jan Schnabel, Lan Cheng, and Andreas K{\"o}hn.
\newblock Limitations of perturbative coupled-cluster approximations for highly
  accurate investigations of {Rb$_2^+$}.
\newblock {\em The Journal of Chemical Physics}, 155(12), 2021.

\bibitem{chakraborty2022benchmarking}
Arnab Chakraborty, Stephen~H Yuwono, J~Emiliano Deustua, Jun Shen, and Piotr
  Piecuch.
\newblock Benchmarking the semi-stochastic {CC (P; Q)} approach for
  singlet--triplet gaps in biradicals.
\newblock {\em The Journal of Chemical Physics}, 157(13), 2022.

\bibitem{lee1989diagnostic}
Timothy~J Lee and Peter~R Taylor.
\newblock A diagnostic for determining the quality of single-reference electron
  correlation methods.
\newblock {\em International Journal of Quantum Chemistry}, 36(S23):199--207,
  1989.

\bibitem{Shen2012}
Jun Shen and Piotr Piecuch.
\newblock Merging active-space and renormalized coupled-cluster methods via the
  cc(p;q) formalism, with benchmark calculations for singlet–triplet gaps in
  biradical systems.
\newblock {\em Journal of Chemical Theory and Computation}, 8(12):4968--4988,
  2012.
\newblock PMID: 26593190.

\bibitem{zhang2022variational}
Yu~Zhang, Lukasz Cincio, Christian~FA Negre, Piotr Czarnik, Patrick~J Coles,
  Petr~M Anisimov, Susan~M Mniszewski, Sergei Tretiak, and Pavel~A Dub.
\newblock Variational quantum eigensolver with reduced circuit complexity.
\newblock {\em npj Quantum Information}, 8(1):96, 2022.

\bibitem{guo2024experimental}
Shaojun Guo, Jinzhao Sun, Haoran Qian, Ming Gong, Yukun Zhang, Fusheng Chen,
  Yangsen Ye, Yulin Wu, Sirui Cao, Kun Liu, et~al.
\newblock Experimental quantum computational chemistry with optimized unitary
  coupled cluster ansatz.
\newblock {\em Nature Physics}, 20(8):1240--1246, 2024.

\bibitem{robledo2025chemistry}
Javier Robledo-Moreno, Mario Motta, Holger Haas, Ali Javadi-Abhari, Petar
  Jurcevic, William Kirby, Simon Martiel, Kunal Sharma, Sandeep Sharma,
  Tomonori Shirakawa, et~al.
\newblock Chemistry beyond the scale of exact diagonalization on a
  quantum-centric supercomputer.
\newblock {\em Science Advances}, 11(25):eadu9991, 2025.

\bibitem{mullinax2025classical}
J~Wayne Mullinax, Panagiotis~G Anastasiou, Jeffrey Larson, Sophia~E Economou,
  and Norm~M Tubman.
\newblock Classical preoptimization approach for adapt-vqe: Maximizing the
  potential of high-performance computing resources to improve quantum
  simulation of chemical applications.
\newblock {\em Journal of Chemical Theory and Computation}, 21(8):4006--4015,
  2025.

\bibitem{JWtransform}
Pascual Jordan and Eugene Wigner.
\newblock {\"U}ber das paulische {\"a}quivalenzverbot.
\newblock {\em Zeitschrift f{\"u}r Physik}, 47(9–10):631--651, 1928.

\bibitem{Ryabinkin2019}
Ilya~G. Ryabinkin, Scott~N. Genin, and Artur~F. Izmaylov.
\newblock Constrained variational quantum eigensolver: Quantum computer search
  engine in the {F}ock space.
\newblock {\em Journal of Chemical Theory and Computation}, 15(1):249--255,
  2019.
\newblock PMID: 30512959.

\bibitem{pople-basis}
Pople~J.A Hariharan~P.C.
\newblock The influence of polarization functions on molecular orbital
  hydrogenation energies.
\newblock {\em Theoretical Chemistry Accounts}, 28:213–222, 9 1973.

\bibitem{Gaussian}
M.~J. Frisch, G.~W. Trucks, H.~B. Schlegel, G.~E. Scuseria, M.~A. Robb, J.~R.
  Cheeseman, G.~Scalmani, V.~Barone, G.~A. Petersson, H.~Nakatsuji, et~al.
\newblock Gaussian 16, revision {C}.01.
\newblock {\em NOTE: it is published by Gaussian, Inc.}, 2016.
\newblock Gaussian, Inc., Wallingford CT.

\bibitem{Terachem1}
I.~S. Ufimtsev and T.~J. Martínez.
\newblock Quantum chemistry on graphical processing units. 3. analytical energy
  gradients and first principles molecular dynamics.
\newblock {\em Journal of Chemical Theory and Computation}, 5:2619--2628, 2009.

\bibitem{Terachem2}
A.~V. Titov, I.~S. Ufimtsev, N.~Luehr, and T.~J. Martínez.
\newblock Generating efficient quantum chemistry codes for novel architectures.
\newblock {\em Journal of Chemical Theory and Computation}, 9:213--221, 2013.

\bibitem{Hay1985}
P.~J. Hay and W.~R. Wadt.
\newblock Ab initio effective core potentials for molecular calculations.
  potentials for {K} to {Au} including the outermost core orbitals.
\newblock {\em Journal of Chemical Physics}, 82(1):299--310, 1985.

\bibitem{Roy2008}
L.~E. Roy, P.~J. Hay, and R.~L. Martin.
\newblock Revised basis sets for the {LANL} effective core potentials.
\newblock {\em Journal of Chemical Theory and Computation}, 4(7):1029--1031,
  2008.

\bibitem{TerachemECP}
C.~Song, L.-P. Wang, and T.~J. Martínez.
\newblock Automated code engine for graphical processing units: Application to
  effective core potential integrals and gradients.
\newblock {\em Journal of Chemical Theory and Computation}, 12:92--106, 2016.

\bibitem{TDDFT}
E.~Runge and E.~K.~U. Gross.
\newblock Density-functional theory for time-dependent systems.
\newblock {\em Physical Review Letters}, 52:997--1000, 1984.

\bibitem{CAMB3LYP}
T.~Yanai, D.~P. Tew, and N.~C. Handy.
\newblock A new hybrid exchange-correlation functional using the
  coulomb-attenuating method (cam-{B3LYP}).
\newblock {\em Chemical Physics Letters}, 393:51--57, 2004.

\bibitem{Schmidt1993}
Michael~W. Schmidt, Kim~K. Baldridge, Jerry~A. Boatz, Steven~T. Elbert, Mark~S.
  Gordon, Jan~H. Jensen, Shiro Koseki, Nikita Matsunaga, Kiet~A. Nguyen,
  Shujun~J. Su, Theresa~L. Windus, Michel Dupuis, and John~A. Montgomery.
\newblock General atomic and molecular electronic structure system.
\newblock {\em Journal of Computational Chemistry}, 14(11):1347--1363, 1993.

\bibitem{Gordon2005}
Mark~S. Gordon and Michael~W. Schmidt.
\newblock Advances in electronic structure theory: Gamess a decade later.
\newblock In Clifford~E. Dykstra, Gernot Frenking, Kwang~S. Kim, and Gustavo~E.
  Scuseria, editors, {\em Theory and Applications of Computational Chemistry:
  The First Forty Years}, pages 1167--1189. Elsevier, Amsterdam, 2005.

\bibitem{Nielsen2005}
Michael~A. Nielsen.
\newblock The fermionic canonical commutation relations and the
  {J}ordan–{W}igner transform.
\newblock School of Physical Sciences, The University of Queensland, Lecture
  Notes, 2005.

\bibitem{IBMscienceadvances}
William~J. Huggins, Kai-Lin Sung, Emilio Cobanera, Michael~J. O'Rourke, Israel
  Klich, Jakob~S. Kottmann, and K.~Birgitta Whaley.
\newblock Grid-based methods for chemistry simulations on a quantum computer.
\newblock {\em Science Advances}, 9(11):eabo7484, 2023.

\bibitem{google2025quantum}
Google~Quantum AI and Collaborators.
\newblock Quantum error correction below the surface code threshold.
\newblock {\em Nature}, 638(8052):920--926, 2025.

\bibitem{javadi2024quantum}
Ali Javadi-Abhari, Matthew Treinish, Kevin Krsulich, Christopher~J Wood, Jake
  Lishman, Julien Gacon, Simon Martiel, Paul~D Nation, Lev~S Bishop, Andrew~W
  Cross, et~al.
\newblock Quantum computing with {Q}iskit.
\newblock {\em arXiv preprint arXiv:2405.08810}, 2024.

\bibitem{wang2015multireference}
Jiaqi Wang, Sivabalan Manivasagam, and Angela~K Wilson.
\newblock Multireference character for 4d transition metal-containing
  molecules.
\newblock {\em Journal of chemical theory and computation}, 11(12):5865--5872,
  2015.

\bibitem{langhoff1974configuration}
Stephen~R Langhoff and Ernest~R Davidson.
\newblock Configuration interaction calculations on the nitrogen molecule.
\newblock {\em International Journal of Quantum Chemistry}, 8(1):61--72, 1974.

\end{thebibliography}

\clearpage
\renewcommand{\thetable}{S\arabic{table}}
\renewcommand{\thefigure}{S\arabic{figure}}

\setcounter{table}{0}
\setcounter{figure}{0}
\section*{SI.1: Raw simulation data}
\label{sec:SI1}
\renewcommand{\thetable}{SI.1-\arabic{table}}
\setcounter{table}{0} 

\begin{table}[htb!]
\centering
\renewcommand{\arraystretch}{1.2}
\setlength{\tabcolsep}{8pt}
\begin{tabular}{l|cccccc}
\toprule
\textbf{Material} 
& \multicolumn{2}{c}{\textbf{HF}} 
& \multicolumn{2}{c}{\textbf{iQCC}} 
& \multicolumn{2}{c}{\textbf{iQCC+PT}} \\
& $\mathbf{S_0}$ (E$_h$) & $\mathbf{T_1}$ (E$_h$)
& $\mathbf{S_0}$ (E$_h$) & $\mathbf{T_1}$ (E$_h$)
& $\mathbf{S_0}$ (E$_h$) & $\mathbf{T_1}$ (E$_h$) \\
\midrule
Q1&-1937.920&-1937.839&-1938.112&-1938.039&-1938.130&-1938.057\\
Q2$^1$&-2648.646&-2648.542&-2648.886&-2648.796&-2648.929&-2648.844\\
Q3&-1951.292&-1951.190&-1951.511&-1951.425&-1951.536&-1951.451\\
Q4$^1$&-2254.544&-2254.440&-2254.826&-2254.733&-2254.876&-2254.789\\
Q5&-1760.482&-1760.378&-1760.726&-1760.635&-1760.737&-1760.653\\
Q6&-1799.516&-1799.414&-1799.739&-1799.651&-1799.765&-1799.681\\
Q7&-1884.420&-1884.309&-1884.635&-1884.530&-1884.661&-1884.564\\
Q8&-1635.312&-1635.220&-1635.543&-1635.457&-1635.567&-1635.484\\
Q9&-1426.756&-1426.641&-1426.986&-1426.884&-1427.003&-1426.907\\
Q10&-1450.134&-1450.033&-1450.318&-1450.223&-1450.334&-1450.238\\
Q11&-1218.182&-1218.083&-1218.421&-1218.325&-1218.444&-1218.354\\
Q12&-1420.817&-1420.697&-1421.035&-1420.930&-1421.064&-1420.962\\
Q13&-1482.926&-1482.805&-1483.125&-1483.014&-1483.148&-1483.043\\
Q14&-1521.946&-1521.841&-1522.136&-1522.035&-1522.160&-1522.057\\
\bottomrule
\end{tabular}
\caption{Hartree-Fock, {iQCC} and {iQCC+PT} energies reported in E$_h$. Materials with 1 were simulated using CAS(100,100) instead of CAS(70,70) due to having over 87 atoms.}
\label{tab:SI-1.1}
\end{table}

\begin{table}[htb!]
\centering
\renewcommand{\arraystretch}{1.2}
\setlength{\tabcolsep}{8pt}
\begin{tabular}{l|cccccc}
\toprule
\textbf{Material} 
& \textbf{HF}
& \textbf{iQCC}
& \textbf{iQCC+PT}
& \textbf{CCSD}
& \textbf{CC-CR(2,3)}
& \textbf{Experiment} \\
\midrule
Q1&2.202&1.982&1.988&1.909&1.872&1.974\\
Q2&2.817&2.453&2.313&1.977&1.892&2.362\\
Q3&2.797&2.361&2.319&2.080&2.014&2.344\\
Q4&2.828&2.535&2.379&2.067&2.000&2.371\\
Q5&2.818&2.449&2.282&2.086&2.012&2.375\\
Q6&2.777&2.390&2.289&2.101&2.036&2.375\\
Q7$^1$&3.011&2.857&2.650&2.482&2.453&2.627\\
Q8$^2$&2.506&2.348&2.256&2.051&1.987&2.202\\
Q9&3.118&2.781&2.612&2.283&2.193&2.520\\
Q10&2.738&2.584&2.614&2.394&2.336&2.485\\
Q11$^2$&2.705&2.594&2.441&2.269&2.150&2.421\\
Q12$^3$&3.274&2.835&2.777&2.472&2.376&2.719\\
Q13&3.289&2.999&2.856&2.729&2.644&2.818\\
Q14$^3$&2.861&2.744&2.809&2.430&2.368&2.818\\
\bottomrule
\end{tabular}
\caption{
Calculated ab initio and Experimental T$_1$ $\rightarrow$ S$_0$ gaps in eV. Experimental values were obtained at 77\,K in toluene except for those noted with superscripts: 1 for THF, 2 for dichloromethane, and 3 for 2-MeTHF \cite{Li2017,Li2020ChemMater}}
\label{tab:SI-1.2}
\end{table}

\begin{table}[htb!]
\centering
\renewcommand{\arraystretch}{1.2}
\setlength{\tabcolsep}{8pt}
\begin{tabular}{l|cccccc}
\toprule
\textbf{Material} 
& \shortstack{\textbf{RO-}\\\textbf{B3LYP}}
& \shortstack{\textbf{RO-}\\\textbf{CAM-B3LYP}}
& \shortstack{\textbf{RO-}\\\textbf{$\omega$PBEh}}
& \shortstack{\textbf{RO-}\\\textbf{$\omega$B97X}}
& \shortstack{\textbf{TD-}\\\textbf{B3LYP}}
& \shortstack{\textbf{TD-}\\\textbf{CAM-B3LYP}}\\
\midrule
Q1 & 1.825 & 2.060 & 2.509 & 2.195 & 2.032 & 2.243 \\
Q2 & 2.257 & 2.515 & 2.507 & 2.654 & 2.491 & 2.689 \\
Q3 & 2.364 & 2.562 & 2.543 & 2.677 & 2.499 & 2.671 \\
Q4 & 2.267 & 2.524 & 2.518 & 2.666 & 2.498 & 2.656 \\
Q5 & 2.368 & 2.564 & 2.543 & 2.678 & 2.568 & 2.757 \\
Q6 & 2.252 & 2.492 & 2.480 & 2.621 & 2.584 & 2.769 \\
Q7 & 2.453 & 2.668 & 2.657 & 2.795 & 2.757 & 2.913 \\
Q8 & 1.974 & 2.287 & 2.308 & 2.454 & 2.240 & 2.424 \\
Q9 & 2.266 & 2.520 & 2.501 & 2.657 & 2.380 & 2.776 \\
Q10 & 2.335 & 2.520 & 2.516 & 2.630 & 2.442 & 2.614 \\
Q11 & 2.249 & 2.470 & 2.450 & 2.595 & 2.444 & 2.775 \\
Q12 & 2.400 & 2.737 & 2.752 & 2.946 & 2.571 & 2.889 \\
Q13 & 2.513 & 2.903 & 2.904 & 3.078 & 2.637 & 2.901 \\
Q14 & 2.164 & 2.422 & 2.413 & 2.800 & 2.698 & 2.910 \\
\bottomrule
\end{tabular}
\caption{
Calculated DFT and TD-DFT T$_1$ $\rightarrow$ S$_0$ gaps reported in eV. TD-DFT calculations used the singlet optimized geometry.
}
\label{tab:SI-1.3}
\end{table}
\clearpage

\section*{SI.2: Additional timing data}
\label{sec:SI2}
\renewcommand{\thetable}{SI.2-\arabic{table}}
\setcounter{table}{0} 
The parallel iQCC quantum solver shows that the ansatz optimization stage is not affected by the number of $P_k$ or the number of qubits due to the small and compact ansatz but is affected by the dressing due to the increase in the number of new terms that are generated during the dressing stage. 
The time spent dressing the 28 entanglers appears to scale polynomially with respect to qubit size, which justifies the approach of trying to use the minimal number of dressings and emphasize solving the longest possible approximate ansatz. 
In Table \ref{tab:SI-2.1}, the number of terms does not hit the predefined term limit (2.5$\times$10$^9$ for CAS(40,40), 3.0$\times$10$^9$ for CAS(50,50), CAS(70,70), and CAS(80,80) and 3.5$\times$10$^9$ for CAS(100,100)) and reflection of \glspl{iQCC} scaling characteristics with a fixed number of entanglers.
Table \ref{tab:Z}, in contrast, better reflects the scaling characteristics and overall timing of \gls{iQCC} when the number of entanglers increases.

Since the truncation of the Hamiltonian has an impact on the RAM storage (due to the memory requirements scaling $\mathcal{O}(n*m)$ with respect to Hamiltonian terms $P_k$ and the number of qubits), Table \ref{tab:SI-2.2} reflects the computational time when the Hamiltonian has reached the term limit. 
The total run time of each iQCC step is far less impacted by the number of qubits and implies that the discrepancy in simulation time is based on the number of Entanglers required to reach the desired ground state. 
If this implementation was entirely optimized on the iQCC+PT energy, then the timings would scale based on how aggressively the Hamiltonian is pruned.

\begin{table}[ht!]
\renewcommand{\arraystretch}{1.2}
\setlength{\tabcolsep}{8pt}
\centering
\begin{tabular}{lcccc}
\toprule
\textbf{CAS} 
& \shortstack{\textbf{Initial}\\ \textbf{Hamiltonian}\\\textbf{terms}}
& \shortstack{\textbf{Final}\\ \textbf{Hamiltonian}\\\textbf{terms}}
& \shortstack{\textbf{ansatz}\\ \textbf{simulation}\\\textbf{time [s]}}
& \shortstack{\textbf{Total}\\ \textbf{dressing}\\\textbf{time [s]}} \\
\midrule
40e,40o & 3,728,045 & 29,237,416 & 1.196 & 4.905 \\
50e,50o & 9,102,387 & 55,421,562 & 1.391 & 10.519 \\
70e,70o & 34,985,635 & 150,723,593 & 1.057 & 29.889 \\
80e,80o & 59,621,497 & 224,377,581 & 1.194 & 48.875 \\
100e,100o & 145,408,245 & 361,577,469 & 1.151 & 98.813 \\
\bottomrule
\end{tabular}
\caption{iQCC timing details for the first iQCC iteration of Q1 singlet ground state in triplet geometry. Each Hamiltonian uses the same number of entanglers (28) with the same order polynomial ansatz (6).}
\label{tab:SI-2.1}
\end{table}

\begin{table}[ht!]
\renewcommand{\arraystretch}{1.2}
\setlength{\tabcolsep}{8pt}
\centering
\begin{tabular}{lccc}
\toprule
\textbf{CAS} 
& \shortstack{\textbf{Ansatz}\\ \textbf{Single Simulation}\\\textbf{Time [s]}}
& \shortstack{\textbf{Total}\\ \textbf{Optimization}\\\textbf{Time [s]}} 
& \shortstack{\textbf{Total}\\ \textbf{Wall Clock Time}\\\textbf{Time [hrs]}} \\
\midrule
40e,40o & 451.38 & 15357.82 & 71.68 \\
50e,50o  & 554.45 & 16098.74 &  87.02 \\
70e,70o & 649.21 & 18199.67 & 107.10 \\
80e,80o & 3584.78 & 82429.61 &  114.84 \\
100e,100o & 10626.89 & 361314.12 & 199.37 \\
\bottomrule
\end{tabular}
\caption{iQCC timing details for the \gls{iQCC} quantum solver of Q1 singlet ground state in triplet geometry. The timing statistics include single Ansatz simulation for the final iteration with first order approximated Ansatz, which represents the time spent computing the energy and the Hessian, total time optimizing the final Ansatz including the multiple optimization steps and bit partitioning, and total \gls{iQCC} quantum solver wall clock time}
\label{tab:SI-2.2}
\end{table}
\newpage
\section*{SI.3 Active Space Orbital Selection}
\label{sec:SI3}
\renewcommand{\thetable}{SI.3-\arabic{table}}
\setcounter{table}{0} 
All Hamiltonians were generated using restricted open-shell Hartree--Fock molecular orbitals. 
The selection criterion was to choose a fixed number of occupied orbitals below the Fermi level and the same number of virtual orbitals above it. 
The active-space orbital selection was guided by the results of the \gls{iQCC}+\gls{PT} and \gls{CCSD} calculations. 
Both methods showed the same qualitative trends when examining how the \gls{CAS} size influences the T$_1\rightarrow$S$_0$ gap, as shown in Table \ref{tab:SI-3.1}.
Due to the substantial increase in simulation time on the compute nodes, CAS(70,70) was selected as a reasonable compromise between accuracy and cost (114 hours per state vs.\ 200 hours), especially since the Hamiltonian-generation time also increases dramatically from CAS(70,70) to CAS(100,100) (approximately 2 hours vs.\ 14 hours).

\begin{table}[htb!]
\centering
\renewcommand{\arraystretch}{1.2}
\setlength{\tabcolsep}{8pt}

\begin{tabular}{r|rrrr}
\toprule
\textbf{CAS} & \textbf{Q1} & \textbf{Q5} & \textbf{Q9} & \textbf{Q13} \\
\midrule

\multicolumn{5}{c}{\textbf{iQCC}} \\
\addlinespace[2pt]
\midrule

\multicolumn{1}{r|}{50}  & 1.999 & 2.632 & 2.820 & 3.095 \\
\multicolumn{1}{r|}{70}  & 1.988 & 2.282 & 2.612 & 2.856 \\
\multicolumn{1}{r|}{100} & 1.932 & 2.316 & 2.841 & 3.153 \\

\midrule
\addlinespace[4pt]
\multicolumn{5}{c}{\textbf{CCSD}} \\
\addlinespace[2pt]
\midrule

\multicolumn{1}{r|}{50}  & 1.922 & 2.577 & 2.630 & 3.013 \\
\multicolumn{1}{r|}{70}  & 1.909 & 2.086 & 2.283 & 2.729 \\
\multicolumn{1}{r|}{100} & 1.698 & 1.999 & 2.770 & 2.965 \\

\bottomrule
\end{tabular}

\caption{T$_1$ $\rightarrow$ S$_0$ gaps as a function of \gls{CAS} selection}
\label{tab:SI-3.1}
\end{table}

\section*{SI.4 Quantum Circuit Equivalents}
\label{sec:SI4}
\renewcommand{\thetable}{SI.4-\arabic{table}}
\setcounter{table}{0} 
\renewcommand{\thefigure}{SI.4-\arabic{figure}}
\setcounter{figure}{0}
The \gls{QCC} ansatz used in the \gls{iQCC} quantum solver is processed at an abstract level, where the entanglers take the general form
\[
T = \exp\!\left( -i\,\tau\, \prod (X,Y,Z,I) / 2 \right).
\]

Entanglers generated by the iQCC method have a well-defined structure:  
they always contain an even number of non-identity Pauli operators and always include a single Pauli-$Y$ acting on the lowest-index qubit.  
In derivative versions of the iQCC formalism, such as ILC entanglers, a non-zero number of Pauli-$Z$ operators may appear; however, in the implementation used here, iQCC entanglers do \textbf{not} include Pauli-$Z$ operators. Each entangler contains exactly 1 Pauli Y operator and an odd number of Pauli X operators. 

These entanglers can be decomposed into executable quantum circuits suitable for quantum hardware, following methods described in the literature \cite{Ryabinkin2019,steiger2025sparsesimulationvqecircuits}. In this work, the inter connectivity of the quantum hardware is not a factor in estimating the number of 2-qubit gates and we assume a hardware that uses CNOT gates instead of other 2-qubit control gates. This means that on well reported QPU architectures \cite{IBMscienceadvances,google2025quantum}, the CNOT gate requirements could be much greater due to a lack of full connectivity.

\begin{figure}[htb!]
    \centering
    \includegraphics[width=0.5\linewidth,trim=4 4 4 8,clip]{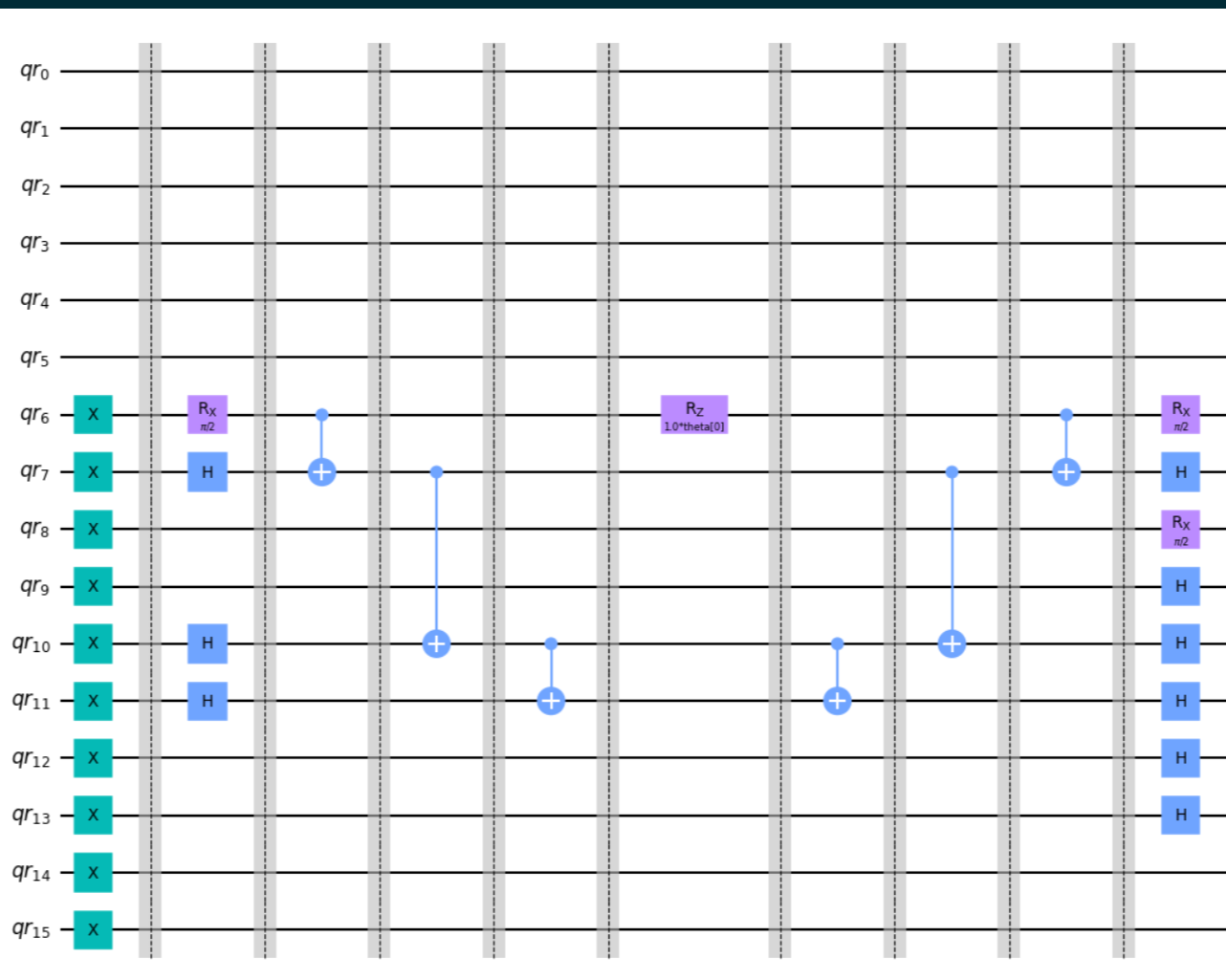}
    \caption{Circuit diagram for diatomic nitrogen generated from QISKIT \cite{javadi2024quantum} }
    \label{fig:SI-4.1}
\end{figure}

Using the first \gls{QCC} entangler for diatomic nitrogen example, the \gls{iQCC} algorithm generates Y(6)*X(7)*X(10)*X(11), which yields an equivalent circuit implemented in QISKIT \cite{javadi2024quantum} shown in Figure~\ref{fig:SI-4.1}.
Using a single CPU, the QISKIT simulator takes 10 hours to sample 812 Hamiltonian terms with 53 entanglers using 4000 samples per Hamiltonian term. 
The \gls{iQCC} quantum solver accomplishes an equivalent task in 0.083 seconds and can optimize the entire set of amplitudes in 0.619028 seconds with the third order Ansatz. 
Furthermore, we show that for diatomic nitrogen, as with water, there is nearly linear speedup efficiency for \gls{iQCC}. 
To achieve the FCI energy (-108.8953259 Ha), we performed 5 iQCC iterations optimizing 10 entanglers for each step followed by 2000 entanglers optimized with the first order approximation, which is accomplished in 90 seconds on a single CPU.

\begin{figure}[htbp!]
    \centering
    \includegraphics[width=0.5\linewidth]{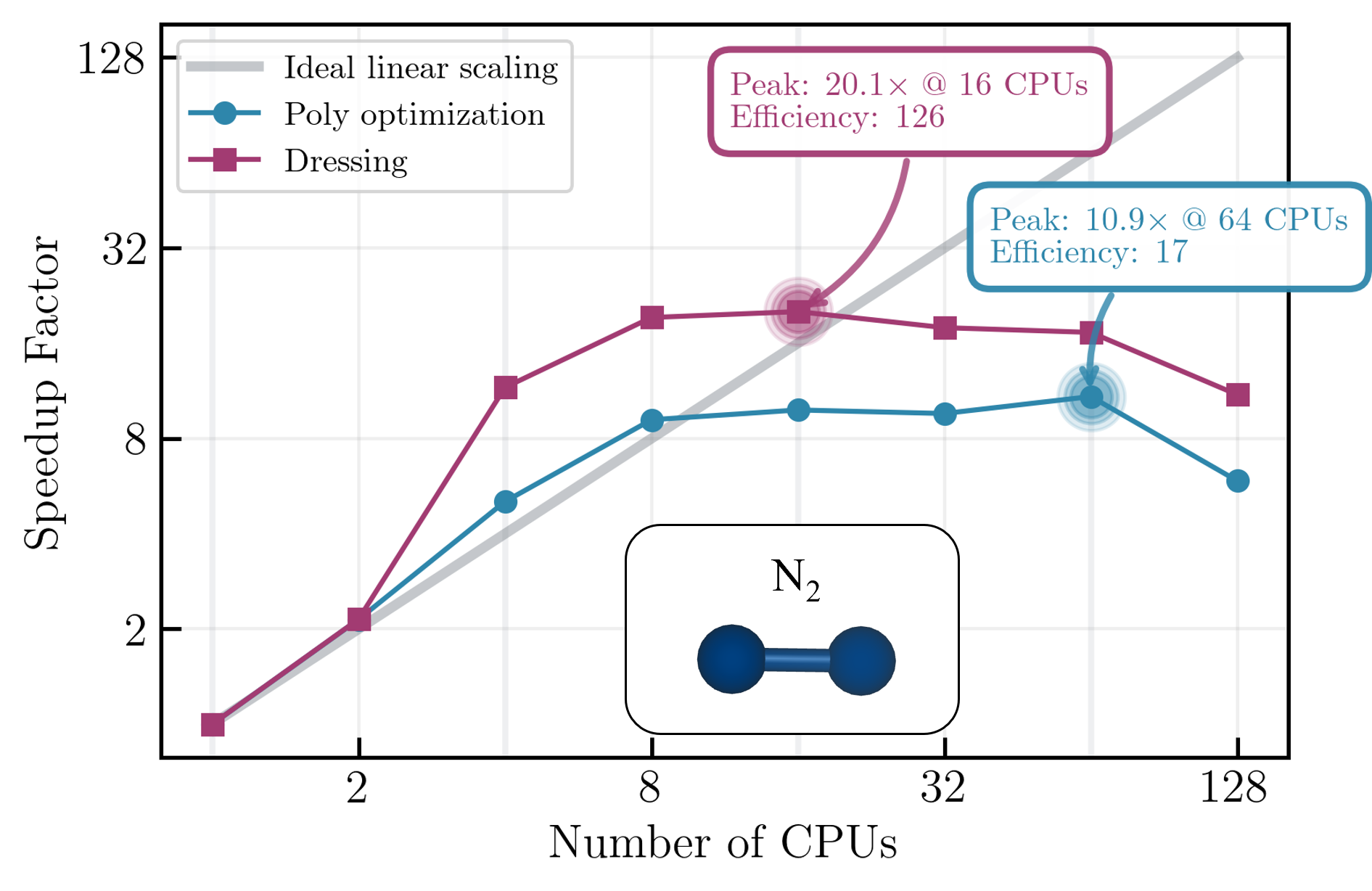}
    \caption{\gls{iQCC} scaling for diatomic nitrogen from 1 to 128 CPUs shows close to linear speedup for both the polynomial optimization and dressing steps of the simulation.}
    \label{fig:SI-4.2}
\end{figure}

The scaling plot of N2 shows the plateau for both the dressing and poly optimization starts at 8 cpus and is at it's peak at 16 cpus Fig.~\ref{fig:SI-4.2}. When considering the scaling for water Fig.~\ref{fig:study}b, this implies the optimal number of cpus is dependent on the system size. CPU scaling studies on Q1 CAS(40,40) on the total runtime, implied that there was not a significant benefit going from 64 processes to 128 processes (total wall clock time of 71.68 hrs and 72.93 hrs respectively).
\newpage
\section*{SI.5: $T1$ and $T2$ Amplitudes from CR-CC(2,3)}
\label{sec:SI5}
\renewcommand{\thetable}{SI.5-\arabic{table}}
\setcounter{table}{0} 

The $T1$ and $T2$ amplitudes from the \gls{CR-CC(2,3)} calculations can be used as a diagnostic tool to estimate the multireference character of a molecule~\cite{lee1989diagnostic}. The $T1$ diagnostic\footnote{This diagnostic is commonly referred to as T$_1$ in the literature but to avoid confusion with the triplet state T$_1$ we define this as $T1$ accordingly.} is defined as the Frobenious norm of the single-excitation amplitude vector normalized by the number of electrons in the molecule, and generally values of $T1 > 0.02 $ are considered to have multireference character~\cite{lee1989diagnostic}, however more recent work suggests that for transition metal organic complexes such as those considered herein, values of $T1 > 0.05$ are more representative~\cite{wang2015multireference}.
Notably, we find that at least for the S$_0$ states, there is very little  predicted multireference character, as all values lie well below both 0.02 and 0.05 benchmarks, as shown in Table \ref{tab:SI-5.1}. 
If we consider the maximum $T1$ and $T2$ values however, we find that the T$_1$ states in most cases have absolute $T1$ and $T2$ values that are  larger than the maximum values in the S$_0$ state, however they are not outside of a reasonable range for a single reference determinant picture, indicating that all diagnostic values suggest this should be a single reference character state.

\begin{table}[htb!]
\centering
\renewcommand{\arraystretch}{1.2}
\setlength{\tabcolsep}{8pt}
\begin{tabular}{l|cc | cc cc}
\toprule
\textbf{Material} 
& \multicolumn{2}{c}{\textbf{$T1$ Diagnostic}} 
& \multicolumn{2}{c}{\textbf{$T1$ Maximum}} 
& \multicolumn{2}{c}{\textbf{$T2$ Maximum}} \\
& \multicolumn{2}{c}{\textbf{$\mathbf{S_0}$}}
& $\mathbf{S_0}$  & $\mathbf{T_1}$ 
& $\mathbf{S_0}$  & $\mathbf{T_1}$ \\
\midrule
Q1 & \multicolumn{2}{c}{0.0112696} & 0.029062 & 0.09158832 & 0.078925 & 0.07431907 \\
Q2 & \multicolumn{2}{c}{0.0094057} & 0.024239 & 0.0705409 & 0.060887 & 0.0462455 \\
Q3 & \multicolumn{2}{c}{0.01013755} & 0.024239 & 0.0705409 & 0.065946 & 0.0647423 \\
Q4 & \multicolumn{2}{c}{0.01051408} & 0.022380 & 0.09688848 & 0.058084 & 0.049424 \\
Q5 & \multicolumn{2}{c}{0.0104161} & 0.020402 & 0.1364138 & 0.060748 & 0.05704999 \\
Q6 & \multicolumn{2}{c}{0.01016506} & 0.023418     & 0.080457     & 0.108029 & 0.110517 \\
Q7 & \multicolumn{2}{c}{0.01296905} & 0.030161     & 0.124574     & 0.073855 & 0.100694 \\
Q8 & \multicolumn{2}{c}{0.00877298} & 0.013624     & 0.071338     & 0.044933 & 0.074148 \\
Q9 & \multicolumn{2}{c}{0.0103547} & 0.020949 & 0.2141291 & 0.046149 & 0.1031473 \\
Q10 & \multicolumn{2}{c}{0.01025959} & 0.022749 & 0.0895289 & 0.045104 & 0.090799 \\
Q11 & \multicolumn{2}{c}{0.01197081} & 0.027926 & 0.1684374 & 0.082651 & 0.0797123 \\
Q12 & \multicolumn{2}{c}{0.00941839} & 0.022673     & 0.145624     & 0.049709 & 0.119305 \\
Q13 & \multicolumn{2}{c}{0.00943724} & 0.023351 & 0.06070263 & 0.061308 & 0.05847821 \\
Q14 & \multicolumn{2}{c}{0.01006642} & 0.030737     & 0.177408     & 0.046328 & 0.039704 \\
\bottomrule
\end{tabular}
\caption{T1 Diagnostic, T1 and T2 Maximum amplitudes in absolute value computed from \gls{CR-CC(2,3)} calculations. Values are extracted from the GAMESS output, where the open-shell T1 diagnostic is not computed, and therefore we include only the T1 diagnostic for singlet states}
\label{tab:SI-5.1}
\end{table}

\section{SI.6 Benchmarking quantum emulation and quantum simulation methods relative to iQCC}
\label{sec:SI-quantum-advantage}
\renewcommand{\thetable}{SI.6-\arabic{table}}
\setcounter{table}{0} 

As shown in Section \nameref{sec:quantum-advantage} of the main text, we benchmark the current implementation of \gls{iQCC} against the leading \gls{VQE} \cite{zhang2022variational,guo2024experimental,Shang2023,mullinax2025classical,Mullinax2025}, and in one case \gls{SQD} \cite{robledo2025chemistry} methods, to-date. The corresponding qubit counts, relative errors, and reference method used for Figure \ref{fig:quantum-advantage} are shown in Table \ref{tab:quantum-advantage}. While \gls{FCI} is the ideal gold-standard reference method, at system sizes larger than a few tens of orbitals, it becomes computationally infeasible to compute the \gls{FCI} reference energies. For this reason, Robledo \textit{et al.} \cite{robledo2025chemistry} chose to use heat bath CI (HCI) as a reference energy, and for our larger molecules we also chose to use \gls{CISD} with a multireference Davidson's correction (CISD+Q) \cite{langhoff1974configuration}. All \gls{FCI} and \gls{CISD}+Q calculations were carried out with GAMESS \cite{Schmidt1993,Gordon2005}. 

\begin{table}[t]
  \centering
  \caption{Benchmarking methods for computing the ground state energy of molecules using quantum emulation or full variational quantum eigensolver models on quantum hardware relative to iQCC. This data is shown in Figure \ref{fig:quantum-advantage} of the main text. }
  \label{tab:quantum-advantage}

  \resizebox{\textwidth}{!}{%
  \begin{tabular}{l l c c c p{2.2cm} p{2.2cm}}
    \toprule

    \multicolumn{1}{c}{\parbox[c]{1.2cm}{\centering Method}} &
    \multicolumn{1}{c}{\parbox[c]{2.0cm}{\centering System}} &
    \multicolumn{1}{c}{\parbox[c]{2.2cm}{\centering Number of atoms}} &
    \multicolumn{1}{c}{\parbox[c]{2.0cm}{\centering Number of Qubits}} &
    \multicolumn{1}{c}{\parbox[c]{2.0cm}{\centering Error per atom /\,$E_h$}} &
    \multicolumn{1}{c}{\parbox[c]{2.0cm}{\centering Reference Method}} &
    \multicolumn{1}{c}{\parbox[c]{2.2cm}{\centering Citation}} \\
    \midrule
ClusterVQE & LiH & 2 & 10 & \num{ 0.00005975} & FCI  & \cite{zhang2022variational} \\
VQE  & F$_2$ & 2 & 12 & \num{ 0.91575} & FCI  & \cite{guo2024experimental} \\
VQE+$\epsilon$ & F$_2$ & 2 & 12 & \num{ 0.0087} & FCI  &  \cite{guo2024experimental} \\
iQCC & N$_2$ & 2 & 16 & \num{0.00001125188} & FCI  & (this work) \\
iQCC+PT & H$_2$O & 3 & 36 & \num{ 0.000732584} & FCI  & (this work)  \\
ADAPT-VQE & H$_2$O & 3 & 46 &  \num{0.0018} & FCI  & \cite{mullinax2025classical} \\
ADAPT-VQE & C$_2$ & 2 & 52 &  \num{0.01335} & FCI  & \cite{mullinax2025classical} \\
IBM SQD & N$_2$ & 2 & 58 &  \num{0.000679058} & HCI  & \cite{robledo2025chemistry} \\
UCCSD & NH$_3$ & 4 & 64 &  \num{0.003953422} & CISD+Q  &  \cite{Mullinax2025} \\
iQCC & NH$_3$ & 4 & 64 &  \num{0.002952278} & CISD+Q  & (this work)  \\
iQCC+PT & NH$_3$ & 4 & 64 &  \num{0.002043022} & CISD+Q  & (this work)  \\
MPS-VQE & H$_2$ & 2 & 92 & \num{ 0.000264} & FCI  & \cite{Shang2023} \\
iQCC & Q1 & 87 & 100 &  \num{0.000287259} & CISD+Q  &(this work)   \\
iQCC+PT & Q1 & 87 & 100 &  \num{0.000201057} & CISD+Q  &(this work)   \\
iQCC & Q1 & 87 & 200 & \num{ 0.000619943} & CISD+Q  &  (this work) \\
iQCC+PT & Q1 & 87 & 200 & \num{ 0.000269782} & CISD+Q  &(this work)   \\
    \bottomrule
  \end{tabular} %
      }
\end{table}

\end{document}